\documentclass[a4paper,fleqn,usenatbib]{mnras}

\usepackage[T1]{fontenc}
\usepackage{ae,aecompl}

\usepackage{geometry}
\usepackage{nicefrac}
\usepackage{float}    
\usepackage{subfig}   
\usepackage{subfloat} 

\usepackage{graphicx}	
\usepackage{amsmath}	
\usepackage{amssymb}	

\newcommand{\mr}[1]{\mathrm{#1}}
\newcommand{\blue}[1]{\textcolor{blue}{#1}}
\newcommand{\red}[1]{\textcolor{red}{#1}}

\newcommand{\mc}[3]{\multicolumn{#1}{#2}{#3}}
\newcommand{\rbpl}{{\em RoboPol}}

\title[The optical polarization of GL and GQ blazars]{\rbpl: The optical polarization of gamma-ray--loud and
  gamma-ray--quiet blazars}

\author[E.~Angelakis et al.]{E.~Angelakis$^{1}$\thanks{E-mail: eangelakis@mpifr.de},
T.~Hovatta$^{2,3}$, 
D.~Blinov$^{4,5,6}$, 
V.~Pavlidou$^{4,6}$,
S.~Kiehlmann$^{2,3}$,
 I.~Myserlis$^{1}$,
\newauthor
M.~B\"{o}ttcher$^{7}$,
P.~Mao$^{8}$,
G.\,V.~Panopoulou$^{4,6}$, 
I.~Liodakis$^{4,6}$,
O.\,G.~King$^{9}$,
M.~Balokovi\'{c}$^{9}$, 
\newauthor
A.~Kus$^{10}$, 
N.~Kylafis$^{4,6}$, 
A.~Mahabal$^{9}$,
A.~Marecki$^{10}$, 
E.~Paleologou$^{4,6}$,
I.~Papadakis$^{4,6}$,
\newauthor
I.~Papamastorakis$^{4,6}$,
E.~Pazderski$^{11}$, 
T.\,J.~Pearson$^{9}$, 
S.~Prabhudesai$^{11}$, 
\newauthor 
A.\,N.~Ramaprakash$^{11}$,
A.\,C.\,S.~Readhead$^{9}$, 
P.~Reig$^{4,6}$,
K.~Tassis$^{4,6}$,
M.~Urry$^{8}$ and
\newauthor 
J.\,A.~Zensus$^{1}$
\\
$^{1}$Max-Planck-Institut f\"{u}r Radioastronomie, Auf dem H\"{u}gel 69, 53121 Bonn, Germany\\
$^{2}$Aalto University Mets\"ahovi Radio Observatory, Mets\"ahovintie 114, 02540 Kylm\"al\"a, Finland\\
$^{3}$Aalto University Department of Radio Science and Engineering,P.O. BOX 13000, FI-00076 AALTO, Finland\\ 
$^{4}$Department of Physics and Institute for Plasma Physics, University of Crete, 71003, Heraklion, Greece\\
$^{5}$Astronomical Institute, St. Petersburg State University,Universitetsky pr. 28, Petrodvoretz, 198504
St. Petersburg, Russia\\
$^{6}$Foundation for Research and Technology - Hellas, IESL, 7110, Heraklion, Greece\\
$^{7}$Potchefstroom Campus Private Bag X6001 Potchefstroom, 2520. South Africa\\ 
$^{8}$Department of Physics, Yale University, New Haven, CT 06520\\
$^{9}$Cahill Center for Astronomy and Astrophysics, California Institute of Technology, 1200 E California
Blvd, MC 249-17, Pasadena CA, 91125,\\ USA\\
$^{10}$Toru\'{n} Centre for Astronomy, Nicolaus Copernicus University, Faculty of
Physics, Astronomy and Informatics, Grudziadzka 5, 87-100 Toru\'{n},\\ Poland\\ 
$^{11}$Inter-University Centre for Astronomy and Astrophysics, Post Bag 4, Ganeshkhind, Pune - 411 007, India
}

\date{Accepted XXX. Received YYY; in original form ZZZ}

\pubyear{2015}

\begin{document}
\label{firstpage}
\pagerange{\pageref{firstpage}--\pageref{lastpage}}
\maketitle

\begin{abstract}
  We present average $R$-band optopolarimetric data, as well as variability parameters, from the first and
  second \rbpl~observing season. We investigate whether gamma-ray--loud and gamma-ray--quiet blazars exhibit
  systematic differences in their optical polarization properties. We find that gamma-ray--loud blazars have a
  systematically higher polarization fraction (0.092) than gamma-ray--quiet blazars (0.031), with the
  hypothesis of the two samples being drawn from the same distribution of polarization fractions being
  rejected at the $3\sigma$ level. {We have not found} any evidence that this discrepancy is related to
  differences in the redshift distribution, rest-frame $R$-band luminosity density, or the source
  classification. The median polarization fraction versus synchrotron-peak-frequency plot shows an envelope
  implying that high synchrotron-peaked sources have a smaller range of median polarization fractions
  concentrated around lower values. Our gamma-ray--quiet sources show similar median polarization fractions
  although they are all low synchrotron-peaked. We also find that the randomness of the polarization angle
  depends on the synchrotron peak frequency. For high synchrotron-peaked sources it tends to concentrate
  around preferred directions while for low synchrotron-peaked sources it is more variable and less likely to
  have a preferred direction.  We propose a scenario which mediates efficient particle acceleration in shocks
  and increases the helical $B$-field component immediately downstream of the shock.
\end{abstract}

\begin{keywords}
galaxies: active -- galaxies: jets -- galaxies: nuclei -- polarization
\end{keywords}



\section{Introduction} 
\label{sec:introduction}


Active galactic nuclei (AGN) are the small fraction of galaxies ($\sim 7\,\%$, \citealt{1995PASA...12..273R})
that appear to have nuclear emission exceeding or comparable to the total stellar output. Of all members of
the AGN class, ``blazars'' are both the most variable sources and the sources that are most common in the
gamma-ray sky \citep{2012ApJS..199...31N,2015ApJS..218...23A}.  With defining characteristic the close
alignment of their confined plasma flow to our line of sight and the often relativistic speeds involved
\citep{1979ApJ...232...34B}, their jet dominates the emission, generally outshining the host galaxy.

Blazars emit radiation throughout the electromagnetic spectrum -- through synchrotron at lower frequencies,
and through inverse Compton, and possibly hadronic processes, at high frequencies. Owing to its synchrotron
character, the blazar jet emission at energies around and below optical frequencies is expected to be
polarized. The polarization levels depend mostly on the degree of uniformity of the magnetic field at the
emission element \citep{1970ranp.book.....P}. The mere detection of some degree of polarization already
implies some degree of uniformity in the magnetic field \citep[e.g.][]{1972Ap&SS..19...25S} and provides a
handle for understanding its topology and strength at the source rest-frame, assuming that the polarized
radiation transmission can be modeled accurately.

In blazars, both the linear polarization degree and angle can show variations over a range of time scales and
magnitudes \citep{1972ApJ...175L...7S,1998AAS...19310714Y,2010PASJ...62...69U}. The polarization angle often
goes through phases of monotonic transition (``rotations'') between two limiting values
\citep{1988A&A...190L...8K}. The detection of such events that specifically appeared to be associated with
episodic activity at high energies \citep{2008Natur.452..966M,2010ApJ...710L.126M,
  2010Natur.463..919A,2014A&A...567A.135A} prompted the use of rotations as a tool to probe the inner regions
of AGN jets and gave rise to a series of different scenarios about the physical processes that may be causing
them.

In order to pursue a systematic investigation of optical polarization properties and the polarization plane
rotations of blazars, we initiated the \rbpl\ high cadence polarization monitoring program
\citep{2014MNRAS.442.1706K,2014MNRAS.442.1693P}. The aim of the program is to study an unbiased subset of a
photon-flux limited sample of gamma-ray--loud (GL) AGN, as well as smaller ``control'' sample of
gamma-ray--quiet (GQ) blazars. The main scientific questions that the program was designed to address are:
\begin{enumerate}
\item Do temporal coincidences between activity at high energies and polarization rotations indeed imply a
  physical connection between the events?
\item What is the temporal polarimetric behavior of blazars?
\item Do the optical polarization properties of GL and GQ blazars differ in a systematic fashion? And are the
  optical polarization and gamma-ray emission independent, or driven by the same process and hence causally
  connected?
\end{enumerate}

First results on the first two questions have been presented in \citet{2015MNRAS.453.1669B} and
\citet{2016MNRAS.457.2252B}. In this paper, we focus on the third question: the optopolarimetric differences
between GL and GQ blazars. On the basis of the exploratory observations conducted during and shortly after the
instrument commissioning, \cite[2013 May -- July,][hereafter: Survey Paper]{2014MNRAS.442.1693P}, we found a
significant difference ($3 \sigma$ level) in the values of the polarization fraction between GL and GQ sources
as measured in a single-epoch survey. The current paper uses data from the first two \rbpl~observing seasons
to verify whether there is indeed a divergence between the two samples and investigate what may be causing it.

The paper is organized as follows: \S \ref{sec:sample_descr} briefly discusses the blazar samples and
observations used in this work. The higher-level data products that we use are presented in \S
\ref{sec:products} along with the maximum likelihood methods used in the estimation of intrinsic mean
values. In \S \ref{sec:analysis} we present a number of studies aiming at investigating the possible
dependence of the polarization on other source properties. In the same section, we test for consistency of
polarimetric properties between GL and GQ sources. Finally, in \S \ref{sec:discussion} we summarize and
discuss our findings within the framework of a shock-in-jet model.

\section{Source Sample and Observations} 
\label{sec:sample_descr}

The details of the source sample selection are discussed in the Survey Paper as well as in
\cite{2015MNRAS.453.1669B}. The GL sample that was monitored during the first two seasons (``main'' GL sample,
62 sources) is a subset of a photon-flux--limited sample of blazars (557 GL sources) from the {\it Fermi}-LAT
Second Source Catalog \cite[2FGL,][]{2012ApJS..199...31N}, to which we applied an $R$-band flux cut as well as
bias-free cuts related to visibility from Skinakas and field quality such as proximity of other field sources.

\cite{2011ApJS..194...29R} have shown that GQ sources have lower radio modulation indices. Therefore, we
select the GQ sample ("main" GQ sample, 15+2 sources) from the source sample of the 15~GHz OVRO monitoring
blazar program \citep{2011ApJS..194...29R} based on radio variability properties and absence from the
2FGL. The selected sources have 15~GHz flux density above 60~mJy and a modulation index higher than
0.02. Additionally, the same $R$-band flux, visibility, and field-quality cuts have been applied to the GQ
sample as to the GL sample. Two of the original GQ sources (c.f. Survey Paper) -- RBPLJ1624+5652 and
RBPLJ1638+5720 -- appear in the Fermi-LAT 4-year Point Source Catalog \citep[3FGL,
][]{2015ApJS..218...23A}. These sources have been replaced by two new control sources.

In Table~\ref{tbl:sample_summary} we list the GL and GQ sources observed at least once during the first two
seasons. In each of the following studies we include any source from that list that satisfies all the
requirements relevant to that study; independently of whether it was monitored or not. The requirements
relevant to each study are stated in the corresponding section.
%
%
%
%
%
%
%
\begin{table*}
\tiny
 \centering
 
 \caption{\label{tbl:sample_summary} 
   Summary of the GL and GQ sources that were observed at least once during the
   first two \rbpl~seasons. For each study we present here we use the subset of the table that satisfies the relevant requirements. Columns: (1) and (7) the \rbpl~ID; (2) and (8) source survey name; (3) and (9) mark
   whether the source is in the TeV \rbpl~or the F-GAMMA
   program; (4) and (10) the 2FGL classification; (5) and (11) source redshift; (6) and (12) number of measurements.}
  \begin{tabular}{llllrrcllllrr}
    \hline
    ID              &Survey ID         &Other$^1$      &Class$^2$ &\mc{1}{c}{z} &N & &ID            &Survey ID        &Other$^1$ &Class$^2$ &\mc{1}{c}{z} &N \\
    (RBPL ...)      &                  &              &          &             & & &(RBPL ...)  &                  &          &      &             & \\
    \hline
    \mc{6}{c}{\bf Monitored Gamma-ray-loud (GL) sources}               &       &               &                  &              &      &      &\\        
    J0045$+$2127 &GB6J0045$+$2127   &              &bzb  &\ldots  & 23      &       &J1121$-$0553 &PKS1118$-$05      &              &bzq   &1.2970 &  1 \\	  
    J0114$+$1325 &GB6J0114$+$1325   &              &bzb  &2.025   & 20      &       &J1132$+$0034 &PKSB1130$+$008    &              &bzb   &0.6780 &  2 \\	  
    J0136$+$4751 &OC457             &F2            &bzq  &0.859   & 24      &       &J1159$+$2914 &Ton599            &F12           &bzq   &0.7250 &  1 \\	  
    J0211$+$1051 &BZBJ0211$+$1051   &              &bzb  &0.2     & 25      &       &J1217$+$3007 &1ES1215$+$303     &TeV F2        &bzb   &0.1300 & 16 \\	  
    J0217$+$0837 &ZS0214$+$083      &              &bzb  &0.085   & 24      &       &J1220$+$0203 &PKS1217$+$02      &              &bzq   &0.2404 &  1 \\	  
    J0259$+$0747 &PKS0256$+$075     &              &bzq  &0.893   & 15      &       &J1221$+$2813 &WComae            &TeV  F12      &bzb   &0.1030 &  7 \\	  
    J0303$-$2407 &PKS0301$-$243     &              &bzb  &0.26    &  6      &       &J1221$+$3010 &PG1218$+$304      &TeV           &bzb   &0.1840 &  2 \\	  
    J0405$-$1308 &PKS0403$-$13      &              &bzq  &0.571   &  5      &       &J1222$+$0413 &4C$+$04.42        &              &bzq   &0.9660 &  1 \\	  
    J0423$-$0120 &PKS0420$-$01      &F12           &bzq  &0.915   &  6      &       &J1224$+$2122 &4C21.35           &TeV  F1       &bzq   &0.4340 &  8 \\	  
    J0841$+$7053 &4C71.07           &F12           &bzq  &2.218   & 13      &       &J1224$+$2436 &MS1221.8$+$2452   &TeV           &bzb   &0.2180 &  5 \\	  
    J0848$+$6606 &GB6J0848$+$6605   &              &bzb  &\ldots  & 14      &       &J1229$+$0203 &3C273             &F12           &bzq   &0.1580 &  1 \\	  
    J0957$+$5522 &4C55.17           &              &bzq  &0.899   &  4      &       &J1230$+$2518 &ON246             &              &bzb   &0.1350 &  1 \\	  
    J0958$+$6533 &S40954$+$65       &F12           &bzb  &0.367   &  9      &       &J1231$+$2847 &B21229$+$29       &              &bzb   &0.2360 &  1 \\	  
    J1037$+$5711 &GB6J1037$+$5711   &              &bzb  &0.8304  & 16      &       &J1238$-$1959 &PMNJ1238$-$1959   &              &agu   &\ldots &  1 \\	  
    J1048$+$7143 &S51044$+$71       &              &bzq  &1.15    &  7      &       &J1245$+$5709 &BZBJ1245$+$5709   &              &bzb   &1.5449 &  1 \\	  
    J1058$+$5628 &TXS1055$+$567     &              &bzb  &0.143   & 12      &       &J1253$+$5301 &S41250$+$53       &              &bzb   &0.1780 &  2 \\	  
    J1203$+$6031 &SBS1200$+$608     &              &bzb  &0.065   & 17      &       &J1256$-$0547 &3C279             &F12           &bzq   &0.5360 & 19 \\	  
    J1248$+$5820 &PG1246$+$586      &              &bzb  &0.8474  & 13      &       &J1314$+$2348 &TXS1312$+$240     &              &bzb   &2.1450 &  1 \\	  
    J1512$-$0905 &PKS1510$-$08      &F12           &bzq  &0.36    & 36      &       &J1337$-$1257 &PKS1335$-$127     &              &bzq   &0.5390 &  1 \\	  
    J1542$+$6129 &GB6J1542$+$6129   &F2            &bzb  &0.117   & 31      &       &J1354$-$1041 &PKS1352$-$104     &F2            &bzq   &0.3320 &  1 \\	  
    J1553$+$1256 &PKS1551$+$130     &F2            &bzq  &1.308   & 30      &       &J1357$+$0128 &BZBJ1357$+$0128   &              &bzb   &0.2187 &  2 \\	  
    J1555$+$1111 &PG1553$+$113      &F1            &bzb  &0.36    & 51      &       &J1427$+$2348 &PKS1424$+$240     &TeV           &bzb   &0.1600 &  7 \\	  
    J1558$+$5625 &TXS1557$+$565     &              &bzb  &0.3     & 34      &       &J1510$-$0543 &PKS1508$-$05      &              &bzq   &1.1850 &  1 \\	  
    J1604$+$5714 &GB6J1604$+$5714   &              &bzq  &0.72    & 25      &       &J1512$+$0203 &PKS1509$+$022     &              &bzq   &0.2190 &  1 \\	  
    J1607$+$1551 &4C15.54           &              &bzb  &0.496   & 25      &       &J1516$+$1932 &PKS1514$+$197     &              &bzb   &1.0700 &  1 \\	  
    J1635$+$3808 &4C38.41           &F12           &bzq  &1.813   & 51      &       &J1548$-$2251 &PMNJ1548$-$2251   &              &bzb   &0.1920 &  1 \\	  
    J1642$+$3948 &3C345             &F12           &bzq  &0.593   & 23      &       &J1550$+$0527 &4C5.64            &              &bzq   &1.4170 &  2 \\	  
    J1653$+$3945 &Mkn501            &F12           &bzb  &0.034   & 52      &       &J1608$+$1029 &4C10.45           &              &bzq   &1.2320 &  2 \\	  
    J1725$+$1152 &1H1720$+$117      &              &bzb  &0.018   & 40      &       &J1637$+$4717 &4C47.44           &              &bzq   &0.7350 &  3 \\	  
    J1748$+$7005 &S41749$+$70       &              &bzb  &0.77    & 45      &       &J1640$+$3946 &NRAO512           &              &bzq   &1.6660 &  1 \\	  
    J1751$+$0939 &OT81              &F2            &bzb  &0.322   & 49      &       &J1643$-$0646 &FRBAJ1643$-$0646  &              &bzb   &\ldots &  1 \\	  
    J1754$+$3212 &BZBJ1754$+$3212   &              &bzb  &\ldots  & 31      &       &J1649$+$5235 &87GB1648$+$5240   &              &bzb   &2.055 & 30 \\        
    J1800$+$7828 &S51803$+$784      &F12           &bzb  &0.68    & 30      &       &J1722$+$1013 &TXS1720$+$102     &              &bzq   &0.7320 &  1 \\	  
    J1806$+$6949 &3C371             &F1            &bzb  &0.05    & 39      &       &J1727$+$4530 &S41726$+$45       &              &bzq   &0.7140 &  1 \\	  
    J1809$+$2041 &RXJ1809.3$+$2041  &              &agu  &\ldots  & 28      &       &J1733$-$1304 &PKS1730$-$13      &F12           &bzq   &0.9020 &  1 \\	  
    J1813$+$3144 &B21811$+$31       &              &bzb  &0.117   & 27      &       &J1745$-$0753 &TXS1742$-$078     &              &bzb   &\ldots &  1 \\	  
    J1836$+$3136 &RXJ1836.2$+$3136  &              &bzb  &\ldots  & 25      &       &J1749$+$4321 &B31747$+$433      &              &bzb   &0.2150 &  1 \\	  
    J1838$+$4802 &GB6J1838$+$4802   &              &bzb  &0.3     & 28      &       &J1813$+$0615 &TXS1811$+$062     &              &bzb   &\ldots &  2 \\	  
    J1841$+$3218 &RXJ1841.7$+$3218  &              &bzb  &\ldots  & 24      &       &J1824$+$5651 &4C56.27           &F1            &bzb   &0.6640 &  2 \\	  
    J1903$+$5540 &TXS1902$+$556     &              &bzb  &\ldots  & 27      &       &J1844$+$5709 &TXS1843$+$571     &              &agu   &\ldots &  1 \\	  
    J1927$+$6117 &S41926$+$61       &              &bzb  &\ldots  & 25      &       &J1848$+$3244 &B21846$+$32B      &              &agu   &\ldots &  1 \\	  
    J1959$+$6508 &1ES1959$+$650     &F1            &bzb  &0.049   & 35      &       &J1849$+$6705 &S41849$+$67       &F2            &bzq   &0.6570 &  1 \\	  
    J2005$+$7752 &S52007$+$77       &              &bzb  &0.342   & 27      &       &J1911$-$1908 &PMNJ1911$-$1908   &              &agu   &\ldots &  1 \\	  
    J2015$-$0137 &PKS2012$-$017     &              &bzb  &\ldots  & 27      &       &J1923$-$2104 &TXS1920$-$211     &F2            &bzq   &0.8740 &  1 \\	  
    J2016$-$0903 &PMNJ2016$-$0903   &              &bzb  &\ldots  & 22      &       &J2000$-$1748 &PKS1958$-$179     &              &bzq   &0.6520 &  1 \\	  
    J2022$+$7611 &S52023$+$760      &              &bzb  &0.594   & 28      &       &J2030$+$1936 &87GB2028$+$1925   &              &agu   &\ldots &  1 \\	  
    J2030$-$0622 &TXS2027$-$065     &              &bzq  &0.671   & 26      &       &J2031$+$1219 &PKS2029$+$121     &              &bzb   &1.2130 &  1 \\	  
    J2039$-$1046 &TXS2036$-$109     &              &bzb  &\ldots  & 32      &       &J2035$+$1056 &PKS2032$+$107     &              &bzq   &0.6010 &  1 \\	  
    J2131$-$0915 &RBS1752           &              &bzb  &0.449   & 28      &       &J2146$-$1525 &PKS2143$-$156     &              &bzq   &0.6980 &  1 \\	  
    J2143$+$1743 &OX169             &F2            &bzq  &0.211   & 29      &       &J2147$+$0929 &PKS2144$+$092     &F2            &bzq   &1.1130 &  1 \\	  
    J2148$+$0657 &4C6.69            &              &bzq  &0.999   & 29      &       &J2152$+$1734 &S32150$+$17       &              &bzb   &0.8740 &  1 \\	  
    J2149$+$0322 &PKSB2147$+$031    &              &bzb  &\ldots  & 23      &       &J2217$+$2421 &B22214$+$24B      &              &bzb   &0.5050 &  1 \\	  
    J2150$-$1410 &TXS2147$-$144     &              &bzb  &0.229   & 20      &       &J2253$+$1404 &BZBJ2253$+$1404   &              &bzb   &0.3270 &  1 \\	  
    J2202$+$4216 &BLLacertae        &F12           &bzb  &0.069   & 77      &       &J2321$+$2732 &4C27.5            &              &bzq   &1.2530 &  1 \\	  
    J2225$-$0457 &3C446             &F1            &bzq  &1.404   & 22      &       &J2325$+$3957 &B32322$+$396      &F2            &bzb   &\ldots &  1 \\	  
    J2232$+$1143 &CTA102            &F12           &bzq  &1.037   & 53      &       & \mc{6}{c}{}                                                     \\	  
    J2243$+$2021 &RGBJ2243$+$203    &              &bzb  &\ldots  & 32      &       & \mc{6}{c}{\bf Monitored Gamma-ray-quiet (GQ) sources}           \\	     
    J2251$+$4030 &BZBJ2251$+$4030   &              &bzb  &0.229   & 33      &       & J0017$+$8135 &\ldots            &                   &RL-FSRQ &\ldots & 11 \\ 	     
    J2253$+$1608 &3C454.3           &F12           &bzq  &0.859   &103      &       & J0642$+$6758 &HB89$-$0636$+$680 &                   &RL-FSRQ &3.1800 & 11 \\	     
    J2311$+$3425 &B22308$+$34       &              &bzq  &1.817   & 30      &       & J0825$+$6157 &HB89$-$0821$+$621 &                   &RL-FSRQ &0.5420 &  8 \\	     
    J2340$+$8015 &TXS2331$+$073     &              &bzq  &0.401   & 13      &       & J0854$+$5757 &HB89$-$0850$+$581 &                   &RL-FSRQ &1.3191 &  6 \\	     
    J2334$+$0736 &BZB J2340$+$8015  &              &bzb  &0.274   & 18      &       & J1551$+$5806 &SBS1550$+$582     &                   &RL-FSRQ &1.3240 & 26 \\	     
    \mc{6}{c}{}                                                             &       & J1603$+$5730 &HB89$-$1602$+$576 &                   &RL-FSRQ &2.8580 & 15 \\	     
    \mc{6}{c}{\bf Not monitored Gamma-ray-loud (GL) sources}                &       & J1624$+$5652 &SBS1623$+$569     &discontinued$^{3}$  &BL~Lac  &0.4150 & 18  \\                                       
    J0136$+$3905 &B30133$+$388      &TeV           &bzb   &\ldots &  4      &       & J1638$+$5720 &HB89$-$1637$+$574 &discontinued$^{3}$  &RL-FSRQ &0.7506 & 24  \\	       
    J0221$+$3556 &S40218$+$35       &F2            &bzq   &0.9440 &  1      &       & J1800$+$3848 &HB89$-$1758$+$388 &                   &RL-FSRQ &2.0920 & 16 \\      	      
    J0222$+$4302 &3C66A             &              &bzb   &0.4440 & 24      &       & J1835$+$3241 &3C382             &                   &\ldots  &0.0579 & 16 \\	  
    J0238$+$1636 &AO0235$+$164      &F12           &bzb   &0.9400 & 19      &       & J1854$+$7351 &S5$-$1856$+$73    &                   &RL-FSRQ &0.4610 & 16 \\	  
    J0340$-$2119 &PKS0338$-$214     &              &bzb   &0.2230 &  1      &       & J1927$+$7358 &HB89$-$1928$+$738 &                   &RL-FSRQ &0.3021 & 13 \\	               
    J0336$+$3218 &NRAO140           &F1            &bzq   &1.2630 &  6      &       & J1955$+$5131 &HB89$-$1954$+$513 &new$^{4}$           &RL-FSRQ &1.2200 &  2  \\	     
    J0339$-$0146 &PKS0336$-$01      &F1            &bzq   &0.8520 &  4      &       & J2016$+$1632 &TXS2013$+$163     &                   &VisS    &\ldots & 11 \\	     
    J0407$+$0742 &TXS0404$+$075     &              &bzq   &1.1330 &  1      &       & J2024$+$1718 &GB6J2024$+$1718   &                   &RL-FSRQ &1.0500 & 13 \\	     
    J0442$-$0017 &PKS0440$-$00      &              &bzq   &0.8450 & 12      &       & J2033$+$2146 &4C$+$21.55        &new$^{4}$           &QSO     &0.1735 &  4  \\	                     
    J0510$+$1800 &PKS0507$+$17      &              &bzq   &0.4160 &  2      &       & J2042$+$7508 &4C$+$74.26        &                   &QSO     &0.1040 & 27 \\	     
    J0721$+$7120 &S50716$+$71       &F12           &bzb   &0.31   & 51      &       & \mc{6}{c}{}                                                     \\	     
    J0738$+$1742 &PKS0735$+$17      &F12           &bzb   &0.4240 & 11      &       & \mc{6}{c}{\bf Not monitored Gamma-ray-quiet (GQ) sources}       \\	     
    J0750$+$1231 &OI280             &F1            &bzq   &0.8890 & 11      &       & J0702$+$8549 &CGRaBSJ0702$+$8549&              &RL-FSRQ     &1.0590 &  1 \\	     
    J0809$+$5218 &1ES0806$+$524     &TeV           &bzb   &0.1370 &  4      &       & J0728$+$5701 &BZQJ0728$+$5701   &              &RL-FSRQ     &0.4260 &  2 \\	     
    J0818$+$4222 &S40814$+$42       &F12           &bzb   &0.5300 & 10      &       & J0837$+$5825 &SBS0833$+$585     &              &RL-FSRQ     &2.1010 &  2 \\	     
    J0830$+$2410 &S30827$+$24       &F1            &bzq   &0.9420 &  6      &       & J1010$+$8250 &8C1003$+$830      &              &RL-FSRQ     &0.3220 &  1 \\	     
    J0854$+$2006 &OJ287             &F12           &bzb   &0.306  & 26      &       & J1017$+$6116 &TXS1013$+$615     &              &RL-FSRQ     &2.8000 &  2 \\	  
    J0956$+$2515 &OK290             &              &bzq   &0.7080 &  1      &       & J1148$+$5924 &NGC3894           &              &BL~Lac - GD &0.0108 &  1 \\     
    J1012$+$0630 &NRAO350           &              &bzb   &0.7270 &  1      &       & J1436$+$6336 &GB6J1436$+$6336   &              &RL-FSRQ     &2.0680 &  1 \\    	    
    J1014$+$2301 &4C23.24           &              &bzq   &0.5650 &  1      &       & J1526$+$6650 &BZQJ1526$+$6650   &              &RL-FSRQ     &3.0200 &  2 \\	  
    J1018$+$3542 &B21015$+$35B      &              &bzq   &1.2280 &  1      &       & J1623$+$6624 &\ldots            &              &RL-FSRQ     &0.201  &  2 \\	  
    J1023$+$3948 &4C40.25           &              &bzq   &1.2540 &  1      &       & J1727$+$5510 &GB6J1727$+$5510   &              &BL~Lac - GD &0.2473 &  4 \\   	    
    J1032$+$3738 &B31029$+$378      &              &bzb   &0.5280 &  3      &       & J1823$+$7938 &S51826$+$79       &              &BL~Lac - GD &0.2240 &  4 \\   
    J1033$+$6051 &S41030$+$61       &              &bzq   &1.4010 &  1      &       & J1850$+$2825 &TXS1848$+$283     &              &RL-FSRQ     &2.5600 &  3 \\   
    J1054$+$2210 &87GB1051$+$2227   &              &bzb   &2.0550 &  1      &       & J1918$+$4937 &BZQJ1918$+$4937   &              &RL-FSRQ     &0.9260 &  3 \\        						    
    J1058$+$0133 &4C1.28            &              &bzb   &0.8880 &  1      &       & J1941$-$0211 &PMNJ1941$-$0212   &              &RL-FSRQ     &0.2020 &  5 \\	     					    
    J1059$-$1134 &PKSB1056$-$113    &              &bzb   &\ldots &  1      &       & J2022$+$6136 &S42021$+$61       &              &RL-FSRQ     &0.2270 &  6 \\	  
    J1104$+$0730 &GB6J1104$+$0730   &              &bzb   &0.6303 &  1      &       & J2051$+$1742 &PKS2049$+$175     &              &Blazar U    &0.1950 &  3 \\	  
    J1104$+$3812 &Mkn421            &F12           &bzb   &0.0300 &  3      &       &     	   &                  &              &            &       &    \\ 
    \hline                                                                        
    \multicolumn{11}{p{12cm}}{$^1$ Indicates whether a source is part of another monitoring sample. ``TeV'' marks
    sources that are in the TeV monitoring sample; ``F'' marks sources of the F-GAMMA sample. The designation
    ``1'' tags F-GAMMA sources before and ``2'' those after F-GAMMA sample change/revision in middle 2009.}\\
    \multicolumn{11}{p{12cm}}{$^2$ Source classification. The tags ``bzq'', ``bzb'' and ``agu'' are taken
    directly from the 2FGL. ``RL-FSRQ''stands for ``QSO RLoud flat radio sp'', ``BL~Lac - GD'' stands for ``BL~Lac - galaxy dominated'' and ``Blazar U'' stand for ``Blazar Uncertain type'' of the Roma BZCAT - 5th edition
    \citep{2015ApSS.357...75M}. Other designations have been taken from NASA/IPAC Extragalactic Database (NED).}\\
    \multicolumn{11}{p{12cm}}{$^3$ discontinued after the completion of the second season.}\\
    \multicolumn{11}{p{12cm}}{$^4$ introduced after the second season (2014) in exchange of the 2 sources that appeared in the 3FGL.}\\
\end{tabular}              
\end{table*}                                                                             

The data sets presented here have been acquired during the first two \rbpl~monitoring seasons, which followed
a brief commissioning phase \citep[2013 May -- July,][]{2014MNRAS.442.1706K,2014MNRAS.442.1693P}. The first
season lasted from 2013 May 26 until 2013 November 27 with 67~per~cent of the observing time usable; the
second season lasted from 2014 April 11 till 2014 November 19 with about 60~per~cent of the nights
usable. Data-taking during each season is discussed in \citet{2015MNRAS.453.1669B} and
\citet{2016MNRAS.457.2252B}, respectively, while our data processing and reduction pipeline is presented in
detail in \cite{2014MNRAS.442.1706K}. The pipeline output includes fractional Stokes parameters $q$
($q=\nicefrac{Q}{I}$) and $u$ ($u=\nicefrac{U}{I}$) and their uncertainties, from which the linear
polarization fraction $p$ and the electric vector position angle (EVPA) $\chi$, for each source are
calculated, with their uncertainties derived from error propagation (see Eqs. 5, 6 in
\citealp{2014MNRAS.442.1706K}).

The median uncertainties of $q$ and $u$ from all measurements in our data set that passed the quality criteria
are both around 0.007 while that of the polarization angle $\chi$, is 4.7\degr. The median uncertainty in
photometry based, for example, on PTF \citep[][]{2012PASP..124..854O} standard stars, is around 0.02~mag. A
measure of the instrumental polarization is given by Table~1 in \cite{2014MNRAS.442.1706K}, where it is shown
that the mean absolute difference between \rbpl-measured and catalogued degree of polarization for polarized
standard stars is about $(3 \pm 5) \times 10^{-2}$ in terms of polarization fraction $p$. Finally, the
instrumental rotation is $2.31\degr\pm0.34\degr$.

After the pipeline operation and before any useful data product is processed, each measurement is subjected to
post-reduction quality checks, which include:
\begin{enumerate}
\item Goodness of the astrometry; by comparing the expected source position to that recovered from the
  reversal of the ``1-to-4'' mapping of the source. The tolerance is 9~arcsec.
\item Field ``crowdedness''; which affects the reliability of the aperture photometry.
\item Central mask edge proximity, which may severely affect the photometry.
\end{enumerate}
All the data products discussed here are based on data sets that have passed all these checks.

\section{Data Products} 
\label{sec:products} 

%
\begin{table*}
 \centering
 \caption{\label{pol_photo_raw_products} 
   Observed polarization parameters and $R$-band magnitudes. $\hat{p}$ is the median
   polarization fraction and $\hat{\chi}$ the median polarization angle. For both quantities only
   measurements that gave significant polarization degree ($\nicefrac{p}{\sigma_p}\ge3$) have
   been used. The angles are corrected for the instrumental rotation. $<R>$ is the mean $R$-band magnitude. No extinction correction has been
   applied to these data. }
  \begin{tabular}{lrrr@{$\pm$}lr@{$\pm$}lr@{$\pm$}lccrr@{$\pm$}lc}
    \hline
      &\mc{10}{c}{Raw polarization parameters}& &\mc{3}{c}{Raw photometric data}\\ 
     \cline{2-11}\cline{13-15} 
    ID           &\mc{1}{c}{N} &\mc{1}{c}{$\left<\tau\right>$} &\mc{2}{c}{$\hat{p}$} &\mc{2}{c}{$p_\mathrm{min}$} &\mc{2}{c}{$p_\mathrm{max}$} &Flag &\mc{1}{c}{$\hat{\chi}$} &  &\mc{2}{c}{$\left<R\right>$} &Photometry $^\mathrm{1}$\\
    (RBPL ...)   &             &(day)                          &\mc{2}{c}{}          &\mc{2}{c}{}               &\mc{2}{c}{}                &    &\mc{1}{c}{($^\circ$)}     & &\mc{2}{c}{(mag)}             &catalogue               \\                                           
    \hline
   J0006$-$0623  &10  &39.2   &0.249   &0.004             &0.104&0.013  &0.355&0.007 &HP &$-$14.5 & &17.21     &0.02   &ST         \\    
   J0035$+$5950  &4   &14.3   &0.033   &0.004             &0.024&0.008  &0.049&0.011 &HP &   81.3 & &17.62     &0.03   &R2         \\ 
   J0045$+$2127  &22  &20.1   &0.078   &0.001             &0.036&0.010  &0.106&0.007 &HP &$-$86.7 & &16.64     &0.01   &ST         \\   
   J0102$+$5824  &14  &18.1   &0.156   &0.004             &0.058&0.017  &0.700&0.104 &HP &$-$82.2 & &17.95     &0.03   &R2         \\ 
   J0114$+$1325  &20  &23.9   &0.066   &0.001             &0.027&0.006  &0.148&0.006 &HP &$-$67.0 & &16.15     &0.01   &PTF        \\ 
   \ldots        &\ldots &\ldots &\mc{2}{c}{\ldots} &\mc{2}{c}{\ldots} &\mc{2}{c}{\ldots} &\mc{1}{c}{\ldots}&\mc{1}{c}{\ldots}& &\mc{2}{c}{\ldots} &\mc{1}{c}{\ldots}  \\
    \hline                                                                                           
   \multicolumn{15}{p{0.85\textwidth}}{$^1$ Label indicating the catalogue used for the absolute photometry
   calibration. ``R2'' is used for USNO$-$B1.0 R2 \citep{2003AJ....125..984M}; ``PTF'' for PTF \citep{2012PASP..124..854O};  ``R1'' for the USNOB1.0 R1 catalogue and
   ``ST'' for photometry based on Landessternwarte Heidelberg-K\"{o}nigstuhl charts.}
  \end{tabular}
\end{table*}                                                                              
 In this section, we present minimal-processing data products for all sources included in
Table~\ref{tbl:sample_summary}.

Table~\ref{pol_photo_raw_products} lists polarimetry and photometry data products for the sources observed.
For polarization angles we adopt the IAU convention: the reference direction is north, and the angle increases
eastwards \citep{1988ARA&A..26...93S}. The Table columns include the number of times $N$ each source has been
observed to be significantly polarized ($\nicefrac{p}{\sigma_p}\ge3$), the average time between two such
consecutive measurements $\langle \tau \rangle$, the median polarization fraction $\hat{p}$, the minimum and
maximum polarization fractions ever observed for each source ($p_{\rm min}$ and $p_{\rm max}$, respectively),
a flag indicating whether the source is of ``high polarization'' (HP) or ``low polarization'' (LP) (with HP
indicating that the source has at some point been observed to have a polarization fraction higher than
$0.03$), and the median polarization angle, $\hat{\chi}$. Polarization angles have been corrected for
instrumental rotation.  The polarization fraction has not been corrected for the host galaxy
  contribution (see Appendix~\ref{sec:host_contr}) or the statistical bias \citep{1974ApJ...194..249W}. We
  consider that the maximum-likelihood data analysis, which we use (see Sect.~\ref{subsec:MLA}), is
  automatically accounting for the statistical bias since only statistically significant values of fractional
  polarization with ($p / \sigma_p \geq 3$) are used, for which the bias is negligible.


Concerning photometry data products, Table~\ref{pol_photo_raw_products} lists the mean $R$-band magnitude for
each source $\langle R \rangle$, averaged over all observations with significant photometry measurements, and
the catalogue used for the photometry calibration.

\subsection{Intrinsic mean flux density and modulation index} 
\label{subsec:MLA_S}
We have used the maximum-likelihood analysis presented in \cite{2011ApJS..194...29R} on the $R$-band flux
densities in order to estimate the intrinsic mean flux density $S_0$ and its modulation index $m_S$, as well
as uncertainties for these quantities. The analysis assumes that, discarding timing information, the
underlying distribution of fluxes that the source is capable of producing is Gaussian. Observational
uncertainties in $R$-band flux density measurements as well as finite sampling are explicitly accounted
for. Table~\ref{tbl:ML_p_S_results} summarizes the results of our analysis.

\subsection{Intrinsic mean polarization and intrinsic modulation index} 
\label{subsec:MLA}

In a similar fashion, we have used a maximum-likelihood analysis to compute best-guess estimates of the
average intrinsic polarization fraction $p_\mr{0}$ and the intrinsic polarization fraction modulation index
$m_p$ ($p-$distribution standard deviation divided by $p-$distribution mean), as well as uncertainties for
these quantities. Physically, $p_\mr{0}$ and $m_p$ correspond to the sample mean and sample modulation index
that one would measure for a source using an infinite number of fair-sampling, zero-observational-error data
points. For this analysis, we have used all measurements, regardless of the signal-to-noise ratio of the
polarization fraction.

The details of the method are described in Appendix A of \cite{2016MNRAS.457.2252B}. The underlying
assumptions are that: (a) a single polarization fraction measurement from a source follows the Rice
distribution (and, implicitly, that the Stokes parameters $Q$ and $U$ have Gaussian, approximately equal
uncertainties); and (b) the values of the polarization fraction that a source can produce follow a Beta
distribution (chosen because it is defined in a closed [0,1] interval, as is the polarization fraction):
\begin{equation}
{\rm PDF}\left(p;\alpha, \beta\right)=\frac{p^{\alpha -1} \left( 1-p\right)^{\beta -1}}{B\left(\alpha, \beta\right)}
\end{equation}
If the parameters $a, \beta$ of this distribution are known, the intrinsic mean and its modulation index are then given by 
\begin{equation}
p_0=\frac{\alpha}{\alpha + \beta}
\end{equation}
and
\begin{equation}
m_p=\frac{\sqrt{\mathrm{Var}}}{p_0}=\frac{\alpha + \beta}{\alpha}\cdot \sqrt{\frac{\alpha\beta}{\left(\alpha + \beta\right)^2 \left(\alpha + \beta
       +1 \right)}}. 
\end{equation}
with Var the variance of the distribution. 

An essential advantage of this approach is that it provides estimates of both uncertainties and, when
appropriate, upper limits. The method has been applied only in cases with at least 3 data points out of which
at least 2 had $\nicefrac{p}{\sigma_p}\ge 3$. All the results of our analysis are shown in
Table~\ref{tbl:ML_p_S_results}.

%
\begin{table*}
\centering
\caption{\label{tbl:ML_p_S_results} 
  Photometry and polarization maximum-likelihood analysis results. $S_0$ is the intrinsic mean $R$-band flux density and $m_S$ its intrinsic modulation index; $p_0$ the mean
  intrinsic polarization fraction and $m_p$ its intrinsic modulation index.}
 \begin{tabular}{lrr@{\hskip .1cm}lrcrr@{\hskip .1cm}lr}
   \hline
      &\mc{4}{c}{maximum-likelihood photometry} & &\mc{4}{c}{maximum-likelihood polarization}\\ 
     \cline{2-5}\cline{7-10} 
   \mc{1}{l}{ID}   &\mc{1}{c}{$S_0$}      & &\mc{1}{c}{$m_S$}      &\mc{1}{c}{N} & &\mc{1}{c}{$p_0$}      & &\mc{1}{c}{$m_p$}      &N\\
   (RBPL ...)      &\mc{1}{c}{(mJy)}     & &                      &             & &                      & &                     & \\
   \hline
   J0006$-$0623 &$0.448^{+0.026}_{-0.026}$ & &$0.175^{+0.055}_{-0.037}$     &11 & &$0.223^{+0.035}_{-0.032}$ & &$0.449^{+0.127}_{-0.091}$     & 11          \\ 
   J0017$+$8135 &$1.169^{+0.017}_{-0.019}$ & &$0.039^{+0.016}_{-0.010}$     &11 & &\ldots                & &\ldots                    &\ldots    \\
   J0035$+$5950 &\dots                 & &\ldots                    &1  & &$0.031^{+0.004}_{-0.005}$ &<&0.642                     &  5           \\ 
   J0045$+$2127 &$0.746^{+0.014}_{-0.014}$ & &$0.087^{+0.016}_{-0.012}$     &23 & &$0.074^{+0.004}_{-0.004}$ & &$0.288^{+0.059}_{-0.048}$     & 23           \\ 
   J0102$+$5824 &$0.194^{+0.021}_{-0.020}$ & &$0.340^{+0.103}_{-0.068}$     &13 & &$0.159^{+0.028}_{-0.022}$ & &$0.520^{+0.133}_{-0.110}$     & 16           \\ 
   \mc{1}{c}{\ldots} &\mc{1}{c}{\ldots}     & &\mc{1}{c}{\ldots}   &\mc{1}{c}{\ldots}  & &\mc{1}{c}{\ldots} & &\mc{1}{c}{\ldots} &\mc{1}{c}{\ldots}\\
   \hline                                                                                                                              
\end{tabular}
\end{table*}                                                                              

\section{Analysis} 
\label{sec:analysis}
Our analysis is focused on the behavior of the polarization fraction $p$ and its variability for GL and GQ
sources. We first examine the median polarization fraction $\hat{p}$ of each source computed from measurements
with $\nicefrac{p}{\sigma_p}\ge 3$. This quantity has the advantage that it is very straightforward to define
and compute. However, it only characterizes sources during their stages of significant polarization, ignoring
non-detections and the associated cycles of low polarization. For this reason, we also include a realistic
analysis which accounts for limited sampling, measurement uncertainties, and Ricean bias, by applying a
maximum-likelihood analysis to compute the intrinsic mean polarization fraction $p_0$ and its associated
intrinsic modulation index $m_p$ (Sect.~\ref{subsec:MLA}), together with uncertainties for these quantities. A
similar approach is followed in for the photometry (Sect.~\ref{subsec:MLA_S}), where a maximum-likelihood
approach is used to compute the intrinsic mean $R$-band flux density $S_0$ and its intrinsic modulation index
$m_S$. The scope of the section can be summarized as (a) quantifying the difference in the amount of
polarization seen on average in GL and GQ sources and its variability, (b) searching for parameters they may
depend on, and (c) investigating the possible scenarios that would explain that difference.

\subsection{The polarization of the GL and GQ samples} 
\label{subsec:PD_GL_GQ}

On the basis of mostly single-measurement data sets collected during the instrument commissioning phase around
2013 May--July, we showed that the polarization fraction of the GL and GQ targets cannot be drawn from the
same parent distributions (see Survey Paper). Assuming an exponential distribution for both classes the mean
values $\left<p\right>$ were $6.4^{+0.9}_{-0.8}\times 10^{-2}$ for GL and $3.2^{+2.0}_{-1.1}\times 10^{-2}$
for GQ sources.

\begin{figure}
\centering
\includegraphics[trim=10pt 0pt 0pt 0pt  ,clip,width=0.49\textwidth]{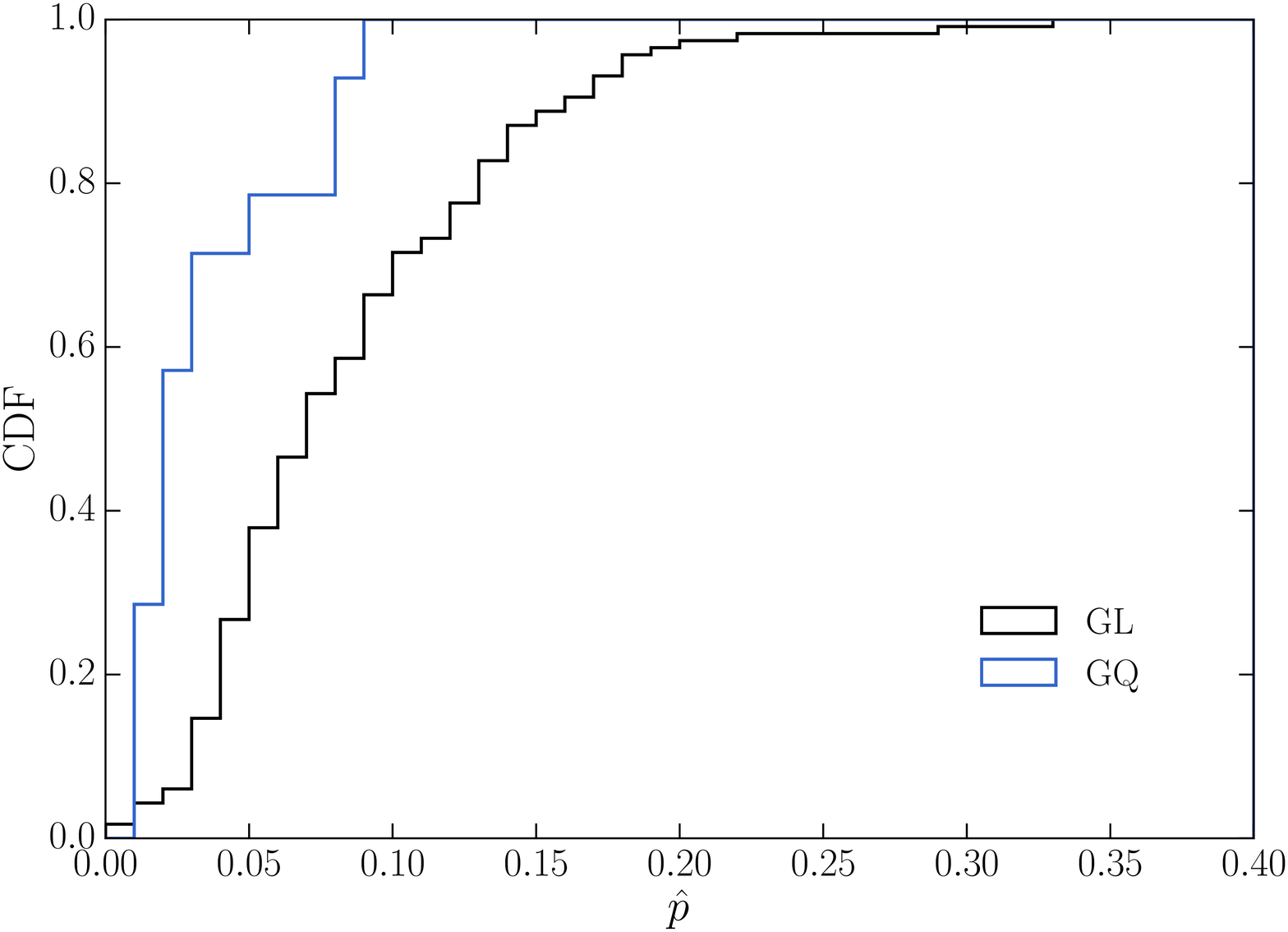} 
\includegraphics[trim=10pt 0pt 0pt 0pt  ,clip,width=0.49\textwidth]{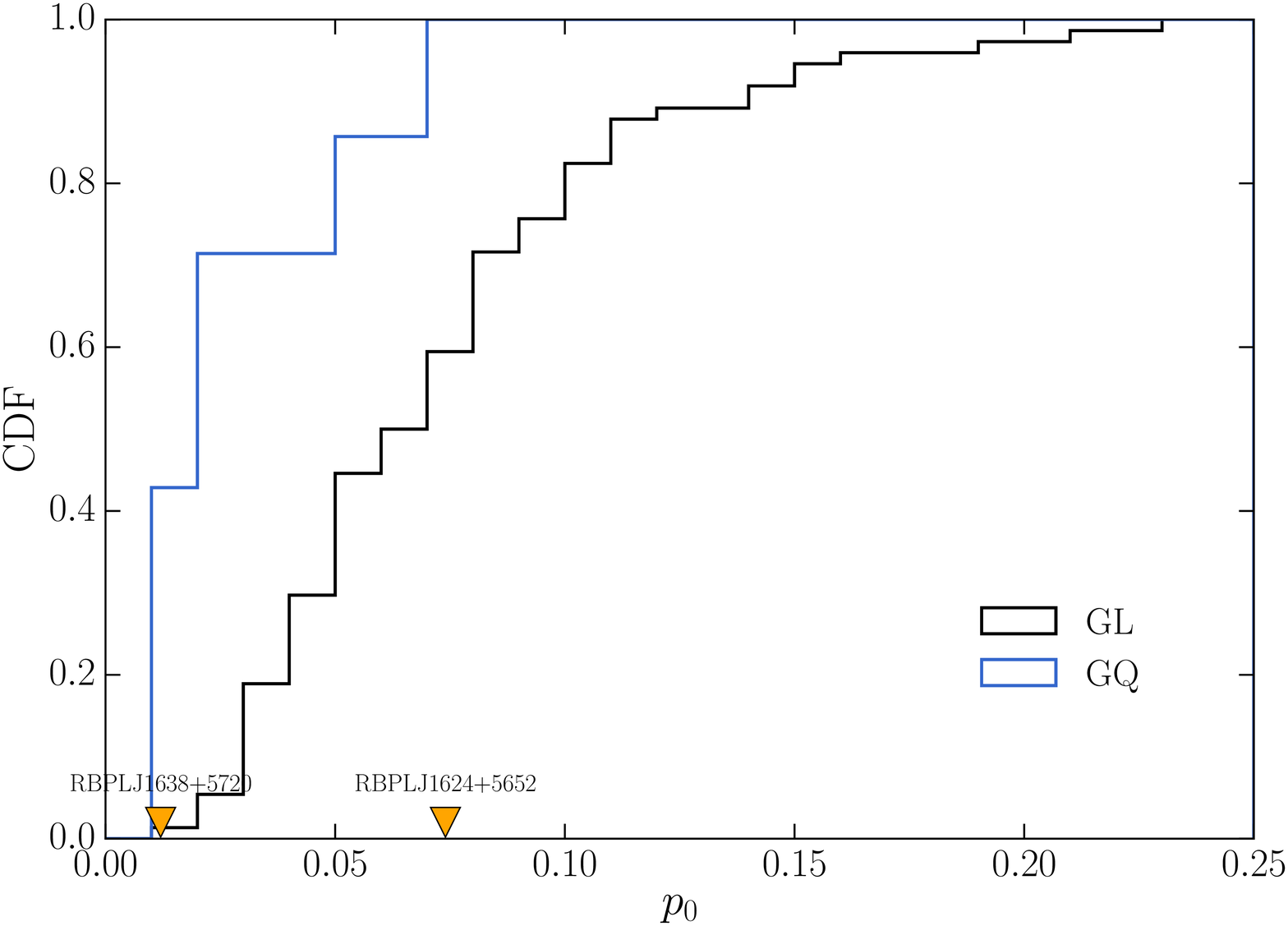} 
\caption{
  The cumulative distribution function of the median polarization fraction for the GL ({\it black}) and GQ
  samples ({\it blue} lines).  {\it Lower:} same for the intrinsic polarization fraction $p_0$. The orange
  triangles indicate the sources that switched from the GQ sample to the GL in the 3FGL catalogue.}
\label{fig:PD-GLoudns}
\end{figure}

Here, we address the same questions using our monitoring data and in particular $\hat{p}$ and $p_0$ for each
source. In the upper panel of Fig.~\ref{fig:PD-GLoudns} we show the cumulative distribution function for the
median polarization fraction $\hat{p}$ of each source. The median is computed from measurements satisfying the
condition $\nicefrac{p}{\sigma_p} \geq 3$. That leaves 116 GL and 14 GQ sources. The median of median
polarization fractions is found to be $0.074\pm0.007$ for the GL sample and $0.025\pm0.009$ for the GQ
ones. The null hypothesis that the two samples come from the same distribution was tested with a two-sample
Kolmogorov-Smirnov (K-S) test which obtained a $D$ of 0.611 and $p$-value of less that $8\times10^{-5}$ (more
than $4\sigma$ significance).

Assuming that $\hat{p}$ follows a log-normal distribution for each sample
\begin{equation}
\mathrm{PDF}=\frac{1}{x\sigma \sqrt{2\pi}} \exp{-\frac{(\ln x - \mu)^2}{2\sigma^2}}
\end{equation}
which would imply an arithmetic mean of 
\begin{equation}
\left<p\right>=e^{\mu + \sigma^2/2}
\end{equation}
and an arithmetic variance of
\begin{equation}
\mathrm{Var}=(e^{\sigma^2}-1) e^{2\mu + \sigma^2}
\end{equation}
we obtain best-fit parameters for the mean $\hat{p}$ and the standard error in the mean
($\sqrt{\nicefrac{\mathrm{Var}}{N}}$ with $N$ the sample length). These are $0.101\pm0.007$ for the GL and
$0.035\pm0.009$ for the GQ samples, respectively.

In the lower panel of Fig.~\ref{fig:PD-GLoudns} we repeat the exercise using the intrinsic polarization
fraction $p_0$ described in Sect.~\ref{subsec:MLA}. There we show 74 GL and 7 GQ sources for which reliable
estimates of $p_0$ have been obtained. The median $\hat{p_0}$ for the two samples is $0.071\pm0.006$ and
$0.020\pm0.011$, respectively. A two sample K-S test gave a p-value of $\sim 2\times10^{-3}$. A major
advantage of the maximum likelihood method is that it provides upper limits. We repeated the previous analysis
including the three GL and the one GQ sources for which only $2\sigma$ upper limits on $p_0$ were
available. We used the nonparametric two-sample tests in the ASURV package \citep{lavalley92}, suitable for
censored data, to estimate the probability that the two distributions come from the same population. According
to Gehan's generalized Wilcoxon test the p-value is $10^{-3}$ indicating the persistence of the difference
between the GL and GQ samples. Assuming again that the two samples are best described by a log-normal
distribution and after including the $2\sigma$ upper limits, the mean intrinsic polarization of the sample
$\left<p_0\right>$ is $0.092\pm0.008$ for GL and $0.031\pm0.008$ for GQ sources. These are the values that we
consider the best-guess to characterize the two source groups.

To examine whether the observed separation is affected by the class of GL sources, we compared the GQ sample
separately with the GL~BL~Lac~objects (sample ``GL-b'') and GL~flat spectrum radio quasars (sample
``GL-q''). Using $\hat{p}$ which is available for larger samples we find that the significance of the
separation remains in the case of GL-b above the $4\sigma$ level while for the GL-q it is around
$2.8\sigma$. We consider the limited size of the latter sample the reason for the lower significance.

To summarize, based on either $\hat{p}$ or $p_0$, GL are on average significantly more polarized than GQ
blazars, and this is not an artifact of different source classes dominating the GL and GQ sample.  In the
following sections we investigate whether this dichotomy can be explained in terms of a dependence on the
redshift, luminosity, the synchrotron peak frequency, color, and source variability.

\subsection{Polarization fraction and redshift} 
\label{subsec:p-z}
In this section, we examine whether $\hat{p}$ shows any dependence on the source redshift, $z$, and whether
the redshift distribution of the members of the GL and GQ samples could be one of the factors responsible for
the different degree of polarization of GL and GQ sources.

\begin{figure}
\centering
\includegraphics[trim=10pt 0pt 0pt 0pt  ,clip,width=0.39\textwidth]{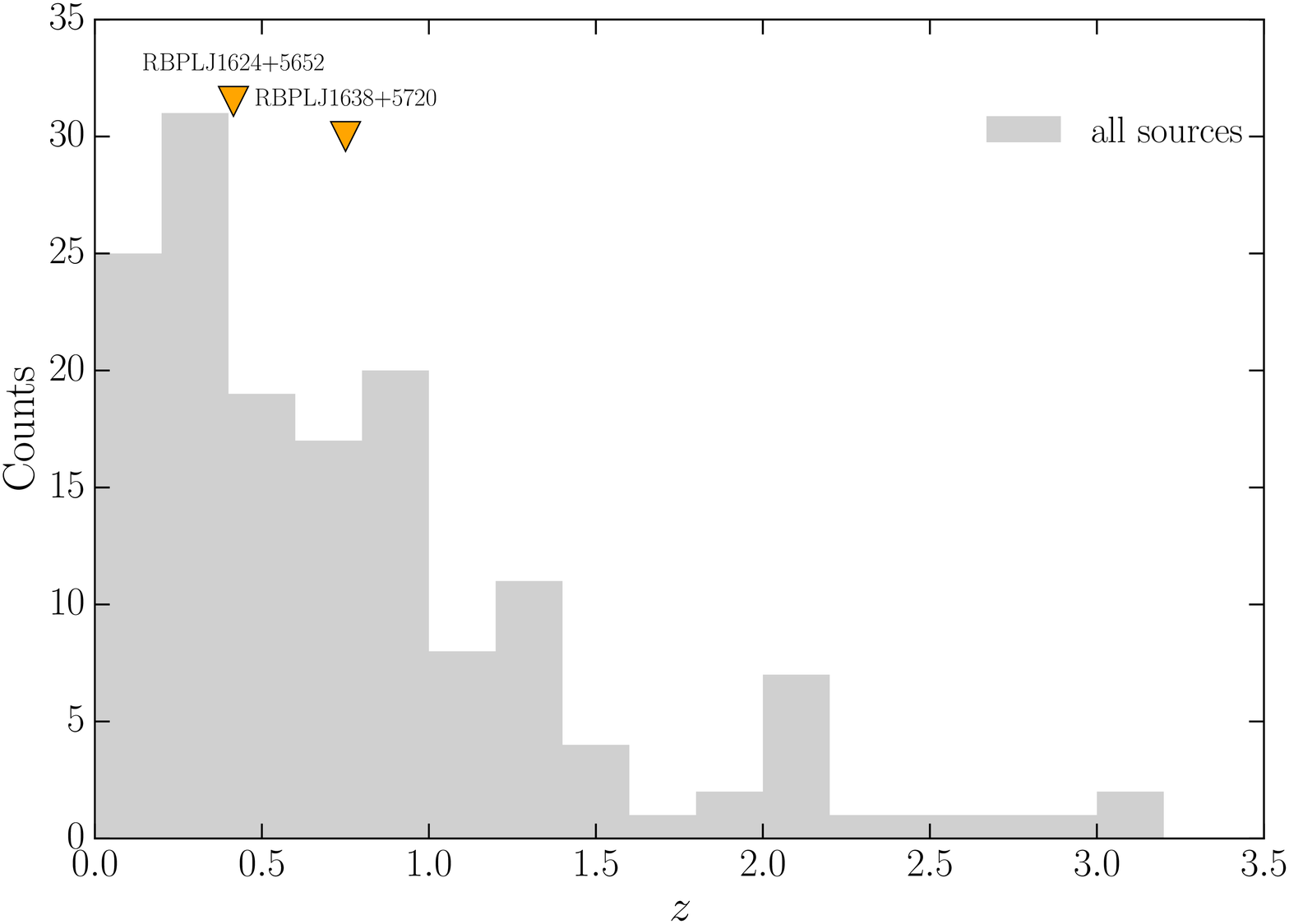} 
\includegraphics[trim=10pt 0pt 0pt 0pt  ,clip,width=0.39\textwidth]{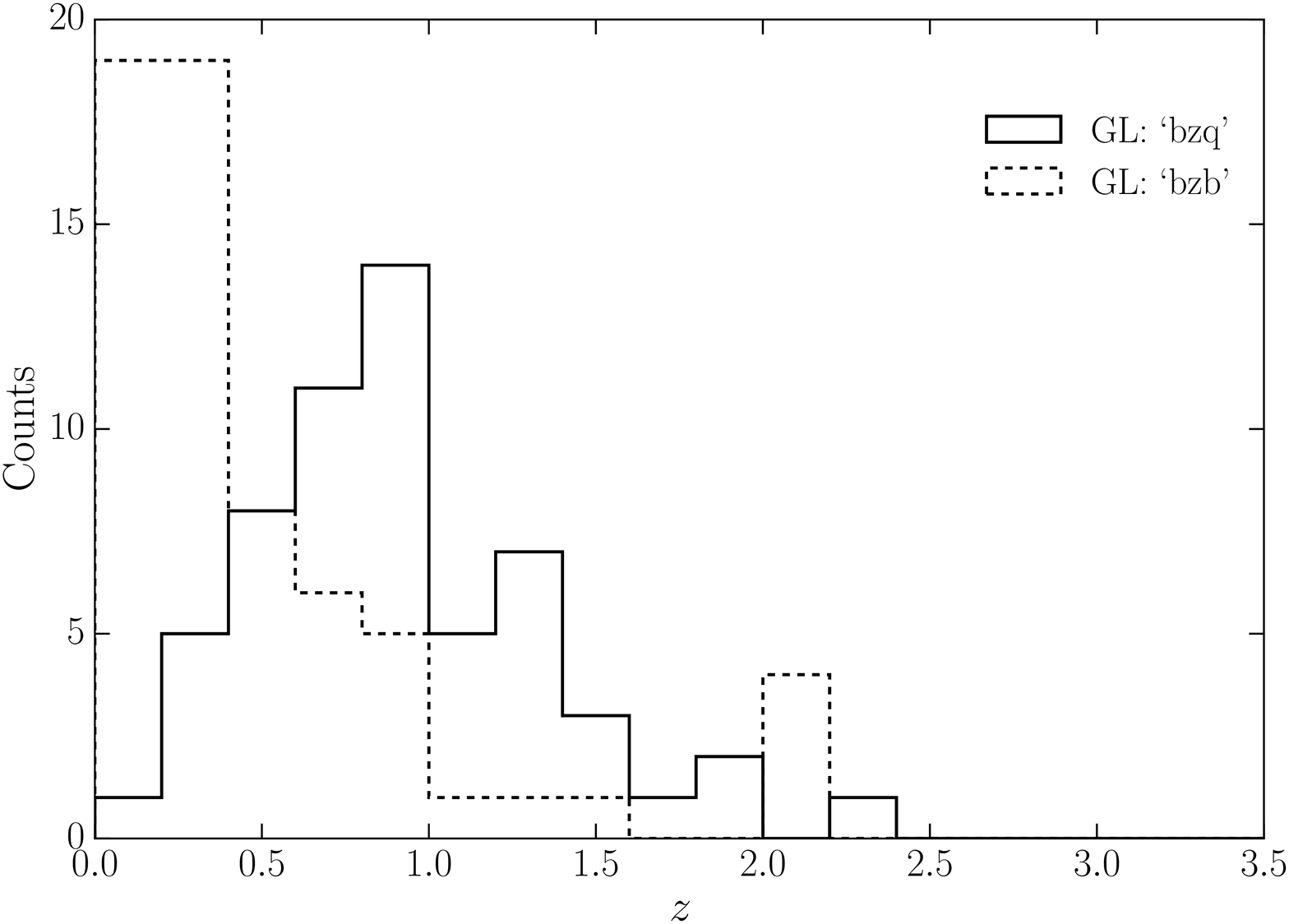} 
\includegraphics[trim=10pt 0pt 0pt 0pt  ,clip,width=0.39\textwidth]{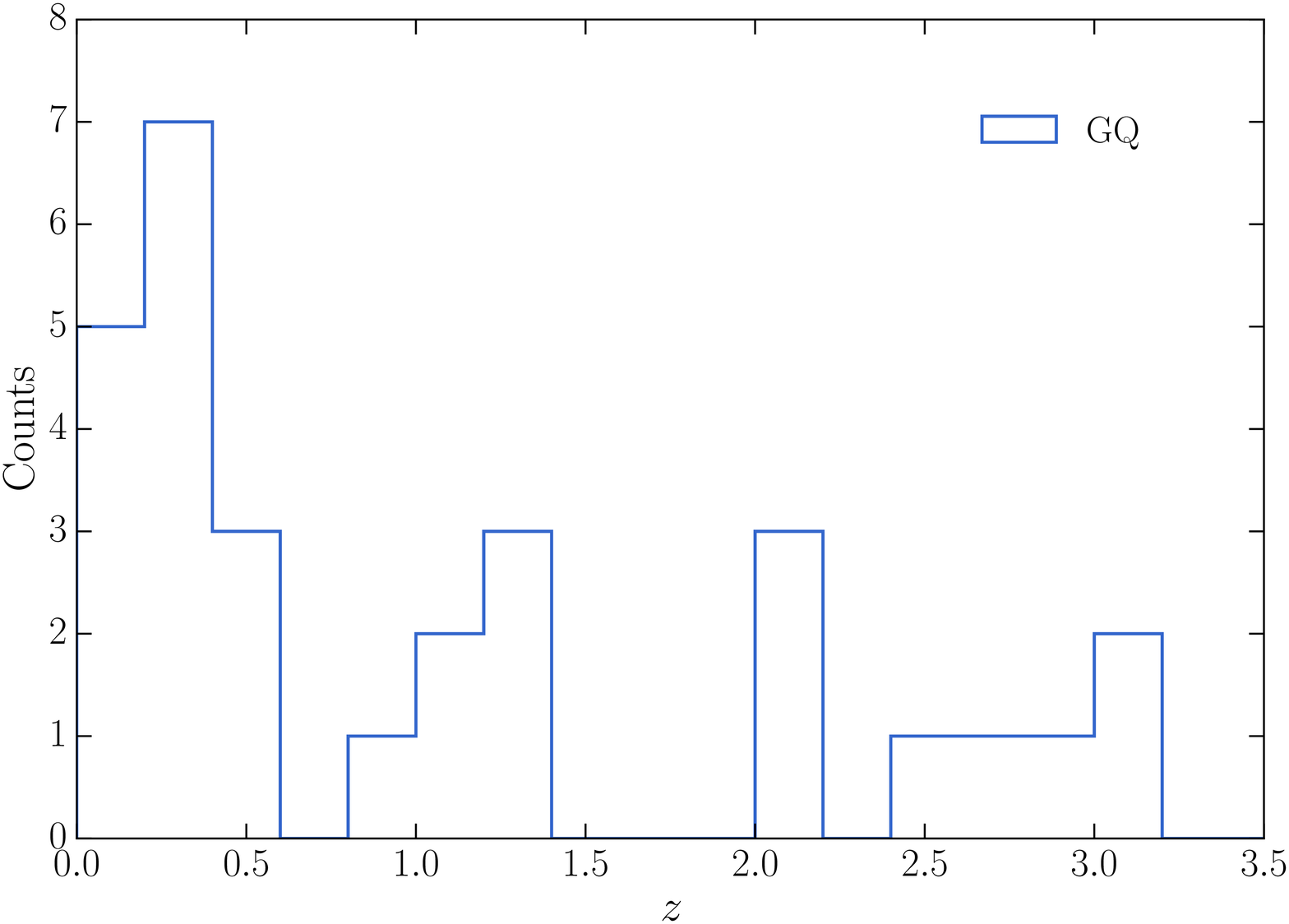} 
\caption{
  The redshift distribution of the main source classes in Table~\ref{tbl:sample_summary}. The bin size is set
  to 0.2. {\it Top panel:} The {\it grey} area shows the distribution of all sources in
  Table~\ref{tbl:sample_summary}. The {\it orange} triangles show the redshifts of the two sources which were
  initially selected as control sample sources but eventually appeared in the 3FGL list. {\it Middle panel:}
  The GL subset is shown separately for ``bzq'' and ``bzb'' sources following the 2FGL classification. {\it
    Bottom panel:} The GQ sources (control sample) is shown in {\it blue}. }
\label{fig:PDF-z}
\end{figure}

In Fig.~\ref{fig:PDF-z} we show the redshift distribution of the GL and GQ sources of our sample. There we
adopt the Roma-BZCAT\footnote{\url{https://heasarc.gsfc.nasa.gov/W3Browse/all/romabzcat.html}} source
designation \citep{2015ApSS.357...75M}: ``bzb'' for BL~Lac objects (i.e. AGNs with a featureless optical
spectrum, or having only absorption lines of galaxian origin and weak and narrow emission lines), and ``bzq''
for flat spectrum radio quasars (with optical spectrum showing broad emission lines and dominant blazar
characteristics). GL sources classified as ``bzb'' are found at systematically lower redshifts (median 0.308)
as opposed to ``bzq'' sources that have a higher median redshift of 0.867, as systematic studies of blazar
samples have shown \citep[e.g.][]{2009A&A...495..691M}. The GQ sources on the other hand are almost uniformly
distributed over a broad range of redshifts reaching up to 3.18. Hence, their cosmological distance cannot
explain -- at least not alone -- their gamma-ray silence. Their median redshift is around 0.5. The orange
triangles mark the positions of the two GQ that appeared in the 3FGL \citep{2015ApJS..218...23A}.

The fact that the quasar subset of blazars (FSRQs) are observed at larger redshifts can impose a mild
dependence of the population admixture on redshift (Fig.~2 in \citealt{2009A&A...495..691M} and Fig.~1 in
\citealt{2015MNRAS.450.3568X}). If at the same time the degree of polarization depended on the source class
(FSRQ or BL~Lac), one could expect an implicit dependence of the polarization fraction on the
redshift. Furthermore, the apparent dominance of quasars in the GQ sample (Table~\ref{tbl:sample_summary})
would impose a similar dichotomy between GL and GQ samples.

As we discuss in Sect.~\ref{subsec:p-L} the contamination of the $R$-band emission by a big blue bump (BBB)
component of thermal origin may modify the intrinsic polarization fraction of a source
\citep[e.g.][]{1993ApJ...415L..83S}. For quasars that are observed at higher cosmological distances this may
become significant. The imbalance of the two main source classes in our samples could naturally introduce
artificial dichotomies. To rule out this possibility we examined the population polarization parameters for
the GL-b and GL-q samples. We found that (a) the two distributions are indistinguishable (K-S test p-value:
0.343), and (b) the mean polarization fraction for the GL-b is $0.087\pm0.005$ and for the GL-q
$0.098\pm0.012$. This excludes the source class as the possible reason for the detected GL--GQ dichotomy.

\begin{figure}
\centering
\includegraphics[trim=10pt 0pt 0pt 0pt  ,clip,width=0.49\textwidth]{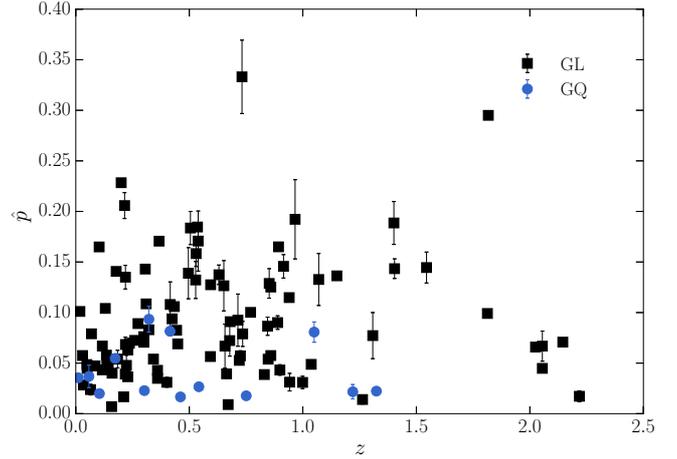} 
\caption{
The median polarization fraction versus the source redshift for GL and
GQ sources. The plot shows no evidence for a monotonic correlation.}
\label{fig:mPD_z.eps}
\end{figure}

Figure~\ref{fig:mPD_z.eps} shows $\hat{p}$ versus $z$ separately for the GL and GQ samples. In order to test
whether $\hat{p}$ depends on $z$ we calculated the Spearman's rank correlation coefficient, $\rho$. The method
assesses the possibility for the existence of a relation between the variables in the form of a monotonic
function. Generally, $\rho$ takes the value of $-1$ or $+1$ in the ideal case of a monotonic relation between
the two variables and 0 in the total absence of such a relation. The case of $\hat{p}$ and $z$ gives a $\rho$
of only 0.18 (p-value: 0.065), lending no support to the hypothesis that there is significant correlation
between the two. The same conclusion is reached when using the intrinsic mean polarization fraction
$p_0$. However, Spearman's test evaluates only the likelihood of a monotonic relation between two variables,
so a more complicated relation cannot be excluded.

Since no strong correlation between redshift and polarization fraction has been identified, we find no
indication that a difference in the redshift distribution between GL and GQ samples can be the source of their
polarization dichotomy.

\subsection{Polarization fraction and luminosity} 
\label{subsec:p-L}
Motivated by the deficiency of apparently bright and highly polarized sources reported in the Survey Paper
(see Fig.~3 therein), we examine the dependence of the source polarization on its $R$-band optical luminosity
density, $L$, and whether such a dependence may be the source of the polarization dichotomy we have identified
between GL and GQ sources.

In the Survey Paper we proposed two alternative explanations for the observed deficiency: (a) the host galaxy
unpolarized starlight contribution \citep[e.g.][]{2008MNRAS.388.1766A} and (b) the dust-induced polarization
\citep[e.g.][]{2015MNRAS.452..715P,2015ARA&A..53..501A} even though at rather low levels ($\sim1$\%). In the
case of AGN blazars this effect must be insignificant as AGNs are generally hosted by dust-poor elliptical
galaxies \citep{2003A&A...400...95N} although not exclusively \citep{1995AJ....110.2027V}.

A third factor that could potentially contaminate the observed emission is that from a BBB
\citep[e.g.][]{1993ApJ...415L..83S}. Depending on its relative intensity, it can contribute unpolarized
emission that may modify the observed polarization fraction. Especially, for quasars this contribution can be
significant and can comprise a considerable fraction of the emission observed in the $R$-band. Under these
circumstances, the observed emission cannot be attributed purely to the jet -- which is our implicit
assumption -- but at least partly to the BBB, as well.

A way around the problem would be to compare jet luminosities, a non-trivial task. Instead, we chose to
investigate the likelihood that our sample suffers from this effect. For 104 sources in
Table~\ref{tbl:sample_summary} SEDs are available from \cite{2016ApJS..224...26M}. Only 15 of these sources
($\sim 14$~\%) showed a clear signature of a BBB, 69 ($\sim 66$~\%) do not have a significant contribution of
a BBB, and for 20 sources the evidence for a BBB is inconclusive. Consequently, the possibility that our
findings are influenced by the contribution of a BBB is negligible. We emphasize that even in cases with clear
contributions of a BBB, the amount of contamination depends on the relative intensity. We conclude that
although the BBB must always be taken into account, its potential contribution to the total intensity in a
small fraction of the \rbpl~sources does not affect our results.

\begin{figure}
\centering
\includegraphics[trim=10pt 0pt 0pt 0pt  ,clip,width=0.49\textwidth]{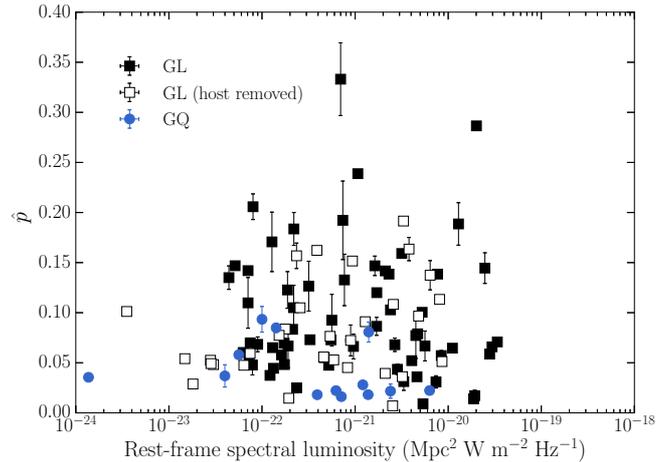} 
\caption{
The median polarization fraction as a function of the $R$-band rest-frame spectral
 luminosity. We show separately the GL (82 squares) and GQ samples (14 circles). For 32 GL and the
 host contribution has been subtracted (empty squares).}
\label{fig:PDvsL_k}
\end{figure}

In Fig.~\ref{fig:PDvsL_k} we show the median polarization fraction $\hat{p}$ as a function of the rest-frame
spectral luminosity for sources in the GL and GQ samples. As we explain in Appendix~\ref{sec:host_contr}, the
luminosity coordinate has been subjected to (a) galactic extinction correction (using extinction values from
the NASA/IPAC Extragalactic Database, NED), (b) host galaxy contribution removal (see
Appendix~\ref{sec:host_contr}), and (c) K-correction assuming a spectral index of $\alpha = -1.3$ for an
optical SED following a power law of the form $\nu^\alpha$
\citep[][]{2004A&A...419...25F,2014MNRAS.439..690H}. In total we show 82 GL and 14 GQ sources. For 32 GL
sources the host galaxy contribution has been removed (c.f. Table~\ref{tab:hosts}). A Spearman's rank-order
correlation coefficient computed for GL and GQ sources collectively gave a correlation index 0.028 (p-value:
0.752), showing no evidence for a monotonic relation. A similar result is found when the host galaxy
contribution is removed from the polarization fraction.

Therefore, there is no indication in our data that the GL-GQ polarization fraction dichotomy can be traced to
a difference in jet luminosity at optical wavelengths between the two samples.

\subsection{Polarization as a function of the synchrotron peak frequency} 
\label{subsec:p-synch_peak}

%
%
\begin{table}
\scriptsize
\centering
\caption{\label{tbl:synch_peaks} 
The logarithm of the rest-frame synchrotron peak frequencies.}
 \begin{tabular}{lrlr}
   \hline
   \mc{1}{c}{ID}   &\mc{1}{c}{Log$\left(\nu_\mathrm{S}/\mathrm{Hz}\right)$} &\mc{1}{c}{ID}   &\mc{1}{c}{Log$\left(\nu_\mathrm{S}/\mathrm{Hz}\right)$}\\
   (RBPL ...)      &                 &(RBPL ...)                &               \\
   \hline
\mc{2}{c}{\bf GL from Mao et al. (2016)}  &\mc{2}{c}{ }\\ 
J0136$+$4751 &13.0           &J1224$+$2122 &13.9      \\ 
J0238$+$1636 &12.9           &J1224$+$2436 &15.4      \\                     
J0259$+$0747 &12.7           &J1229$+$0203 &13.5      \\                     
J0423$-$0120 &12.7           &J1230$+$2518 &14.9      \\               
J0442$-$0017 &13.0           &J1231$+$2847 &15.0      \\               
J0510$+$1800 &13.1           &J1238$-$1959 &14.1      \\               
J0750$+$1231 &13.1           &J1245$+$5709 &14.8      \\               
J0841$+$7053 &12.5           &J1248$+$5820 &14.9      \\               
J0957$+$5522 &13.0           &J1253$+$5301 &13.9      \\               
J0958$+$6533 &13.2           &J1314$+$2348 &14.9      \\               
J1159$+$2914 &13.3           &J1357$+$0128 &14.8      \\               
J1222$+$0413 &14.0           &J1427$+$2348 &15.3      \\               
J1256$-$0547 &13.0           &J1512$+$0203 &13.6      \\               
J1337$-$1257 &13.0           &J1516$+$1932 &13.0      \\               
J1512$-$0905 &13.3           &J1542$+$6129 &14.6      \\               
J1553$+$1256 &13.0           &J1555$+$1111 &15.5      \\               
J1604$+$5714 &13.1           &J1558$+$5625 &14.2      \\               
J1635$+$3808 &12.7           &J1607$+$1551 &13.4      \\               
J1637$+$4717 &12.8           &J1649$+$5235 &14.4      \\               
J1642$+$3948 &12.7           &J1653$+$3945 &16.1      \\               
J1722$+$1013 &12.8           &J1725$+$1152 &16.0      \\               
J1751$+$0939 &12.7           &J1727$+$4530 &13.2      \\               
J1800$+$7828 &13.5           &J1748$+$7005 &13.8      \\               
J1824$+$5651 &12.9           &J1749$+$4321 &13.2      \\               
J1849$+$6705 &13.0           &J1754$+$3212 &14.3      \\               
J2000$-$1748 &12.4           &J1806$+$6949 &14.7      \\               
J2005$+$7752 &13.4           &J1809$+$2041 &15.4      \\               
J2143$+$1743 &14.1           &J1813$+$0615 &14.1      \\               
J2148$+$0657 &13.2           &J1813$+$3144 &15.0      \\               
J2225$-$0457 &12.5           &J1836$+$3136 &14.9      \\               
J2253$+$1608 &13.2           &J1838$+$4802 &15.8      \\               
J2311$+$3425 &13.0           &J1841$+$3218 &16.3      \\               
J2334$+$0736 &12.8           &J1844$+$5709 &14.3      \\               
\mc{2}{c}{ }		     &J1903$+$5540 &14.4      \\               
\mc{2}{c}{\bf GL from 3FGL}  &J1911$-$1908 &15.9      \\               
J0045$+$2127 &16.0           &J1927$+$6117 &13.4      \\               
J0114$+$1325 &15.0           &J1959$+$6508 &16.9      \\               
J0136$+$3905 &16.2           &J2015$-$0137 &14.4      \\               
J0211$+$1051 &14.1           &J2022$+$7611 &14.1      \\               
J0217$+$0837 &13.8           &J2030$-$0622 &13.2      \\               
J0222$+$4302 &15.1           &J2030$+$1936 &15.6      \\               
J0303$-$2407 &15.4           &J2039$-$1046 &13.8      \\               
J0336$+$3218 &13.4           &J2131$-$0915 &16.8      \\               
J0339$-$0146 &13.1           &J2149$+$0322 &14.1      \\               
J0340$-$2119 &13.5           &J2150$-$1410 &17.1      \\               
J0721$+$7120 &14.0           &J2202$+$4216 &13.6      \\               
J0738$+$1742 &14.0           &J2217$+$2421 &13.4      \\               
J0809$+$5218 &15.9           &J2232$+$1143 &12.7      \\               
J0818$+$4222 &13.0           &J2243$+$2021 &15.6      \\               
J0830$+$2410 &12.8           &J2251$+$4030 &14.6      \\               
J0848$+$6606 &14.7           &J2340$+$8015 &15.6      \\               
J0854$+$2006 &13.7           &\mc{2}{c}{ }            \\	                 
J1032$+$3738 &14.1           &\mc{2}{c}{\bf GQ from Mao et al. (2016)}\\             
J1033$+$6051 &13.5           &J0825$+$6157 &12.7        \\             
J1037$+$5711 &14.7           &J1551$+$5806 &13.8        \\             
J1048$+$7143 &13.2           &J1638$+$5720 &12.8        \\             
J1054$+$2210 &14.6           &J1854$+$7351 &13.4        \\             
J1058$+$5628 &15.1           &J1955$+$5131 &13.2        \\             
J1059$-$1134 &13.6           &J2024$+$1718 &13.4        \\             
J1104$+$0730 &14.6           &\mc{2}{c}{ }		 \\                
J1104$+$3812 &17.1           &\mc{2}{c}{\bf GQ from 3FGL}\\            
J1132$+$0034 &14.1           &J1624$+$5652 &13.6         \\            
J1203$+$6031 &14.9           &\mc{2}{c}{ }		   \\              
J1217$+$3007 &15.3           &\mc{2}{c}{\bf GQ from Lister et al. (2015)} \\                      
J1221$+$2813 &14.4           &J1927$+$7358 &13.2          \\           
J1221$+$3010 &16.7           &\mc{2}{c}{ }                \\
   \hline                                                         
\end{tabular}
\end{table}                                                                              

\begin{figure}
\centering
\includegraphics[trim=10pt 0pt 0pt 0pt  ,clip,width=0.49\textwidth]{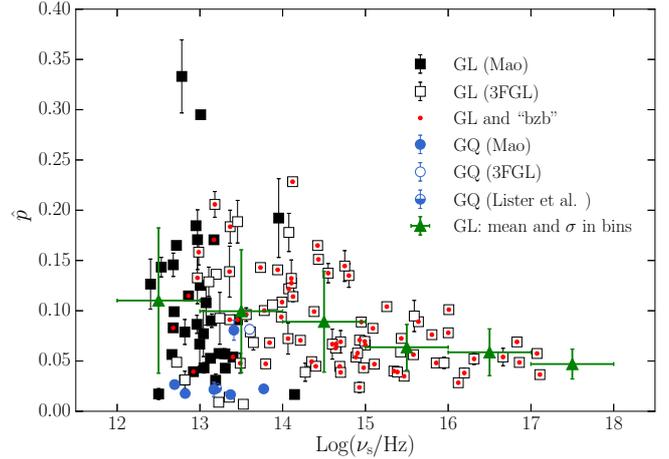} 
\caption{
  The polarization fraction as a function of the rest-frame synchrotron peak frequency. The squares mark GL
  sources and the circles GQ ones. For the filled symbols the peak frequency taken from Mao et al. (2016)
  while for the empty ones from 3FGL or Lister et al. (2015). The {\it red} dots denote the BL~Lac subset of
  GL sources. The {\it green} triangles correspond to the mean within each frequency bin. The bin width is
  marked with the $x$-axis error-bar and has a total length of one. The $y$-axis error-bars have a length of
  one standard deviation computed within the bin.}
\label{fig:PD_peak}
\end{figure}

The location of the synchrotron peak may be another factor affecting the average polarization properties of
the GL and GQ samples. To study such a possible effect we plot, in Fig.~\ref{fig:PD_peak}, the median
polarization fraction $\hat{p}$ against the logarithm of rest-frame synchrotron peak frequency for the GL and
GQ sources.

The synchrotron peak frequencies -- for both samples -- were estimated through a second order polynomial fit
to the synchrotron peak of their SED using data presented in \cite{2016ApJS..224...26M}. Their datasets
include two radio frequencies -- at 1.4 \citep[from NVSS and FIRST catalogs][]{Condon1998AJ,White1997ApJ} and
at 5~GHz from the GB6 and PMN catalogs \citep{Gregory1996ApJS,Wright1994ApJS} -- four infrared frequencies
from WISE,\footnote{\url{http://wise2.ipac.caltech.edu/docs/release/allwise/}} and four optical filters (z, i,
r, g) from the SDSS DR9 \citep{ahn12}. The X-rays were extracted from the Swift archive \citep{burrows05} and
gamma rays from the 3FGL \citep{2015ApJS..218...23A}.  All fluxes were K-corrected to the rest frame before
obtaining luminosities and the fitting was done in logarithmic luminosity space. Details of the data and the
corrections applied to them are given in \cite{2016ApJS..224...26M}.  Although the SDSS u-band has been
excluded from our dataset to avoid the influence of a possible BBB, such a contribution may still be
present. For that reason we inspected all our SEDs to identify problematic cases. Indeed, for 16 of the GL
sources we found that a BBB had or could have had an effect on the localization of the peak. For those cases
the synchrotron peak frequency was taken from the 3FGL \cite{2015ApJ...810...14A}, instead. For the GQ sample,
three sources could have been affected by the presence of a BBB. For two of them peak values were available
from the 3FGL and \cite{2015ApJ...810L...9L} while the third one was excluded. All the values used here are
given in Table~\ref{tbl:synch_peaks}.

Figure~\ref{fig:PD_peak} shows the dependence of median polarization on synchrotron peak frequency. It can be
seen there that there is an upper envelope that decreases with increasing synchrotron peak frequency. However,
a Spearman test does not favor a significant monotonic anti-correlation. The anti-correlation strength is only
$\rho=-0.2$ (p-value: 0.04), when calculated collectively for all GL and GQ sources. For the GL sources,
however, the synchrotron peak frequency estimates are more reliable owing to the better and denser datasets
available. Applying the test to the GL sources alone revealed some anti-correlation with a $\rho$ around
$-0.3$ and a p-value of $2\times10^{-3}$. If the test is further restricted to only the BL~Lac subset of GL
sources (classified as ``bzb'') which happen to cover a larger range of peak frequencies, it yields a
$\rho\approx-0.5$ and a p-value of $6\times10^{-6}$.

In Fig.~~\ref{fig:PD_peak} we also plot the mean $\hat{p}$ ({\it green} markers) in each bin. The abscissa
error bars mark the bin extent while the ordinate error bars show the spread of the $\hat{p}$ within the bin
(error bar size is 1 standard deviation). A linear fit to the bin means gives a significant slope of
$-0.012\pm0.001$.

It is clear then that low-synchrotron peaked (LSP) sources appear more polarized than high-synchrotron peaked
(HSP) ones (with LSP, if: $\mathrm{log}(\nu_\mathrm{s})<14$, ISP if: $14\le\mathrm{log}(\nu_\mathrm{s})<15$
and HSP if: $\mathrm{log}(\nu_\mathrm{s})\ge15$, respectively); at the same time their polarization varies
over a broader range. However, as GQ sources are preferentially LSPs, this trend cannot explain their
systematically lower polarization compared to GL sources.

\subsection{Polarization angle randomness as a function of the synchrotron peak frequency} 
\label{subsec:chi2_peak}
\begin{figure}
\centering
  \includegraphics[trim=10pt 0pt 0pt 0pt  ,clip,width=0.48\textwidth]{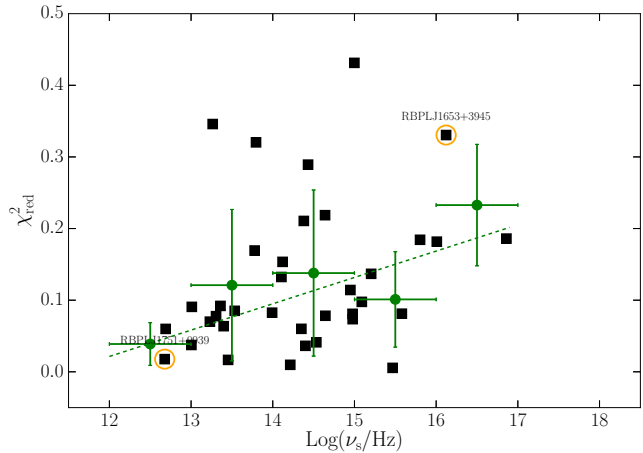} 
\caption{
  The randomness of EVPA as a function of the logarithm of the synchrotron peak frequency. The $y$-axis is the
  reduced $\chi^2$ of the comparison of the source angle distribution with a uniform one. The {\it green}
  solid circles mark the average $\chi^2_\mathrm{red}$ in 5 bins. Their $x$-axis error-bars have a length of
  half a bin-width, while the $y$-axis mark the spread of $\chi^2_\mathrm{red}$ (one standard deviation)
  within that bin. The {\it green} dashed line is the best fit to the binned data (green points).  The {\it
    orange} circles mark one case of high randomness of the EVPA (i.e. close to uniform), RBPLJ1751+0939, and
  one case with low randomness (i.e. far from uniform), RBPLJ1653$+$3945. The angle distributions of these two
  cases are shown in Fig.~\ref{fig:chi_peak_examples}.}
\label{fig:chi_peak}
\end{figure}
\begin{figure*}
\centering
\begin{tabular}{@{\hskip 0.02cm}c@{\hskip 0.02cm}c}
  \includegraphics[trim=10pt 0pt 0pt 0pt  ,clip,width=0.4\textwidth]{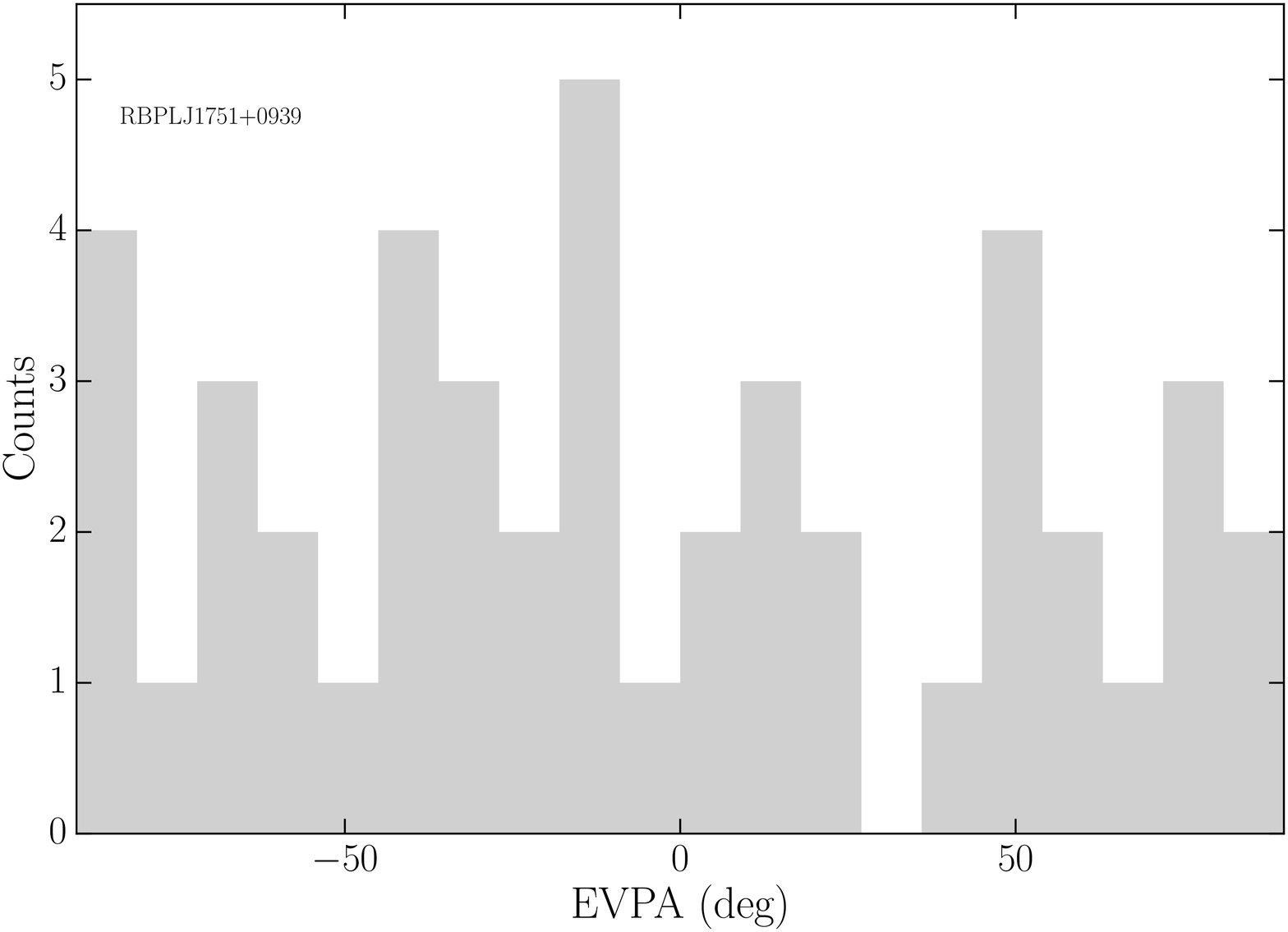}&\includegraphics[trim=10pt 0pt 0pt 0pt  ,clip,width=0.4\textwidth]{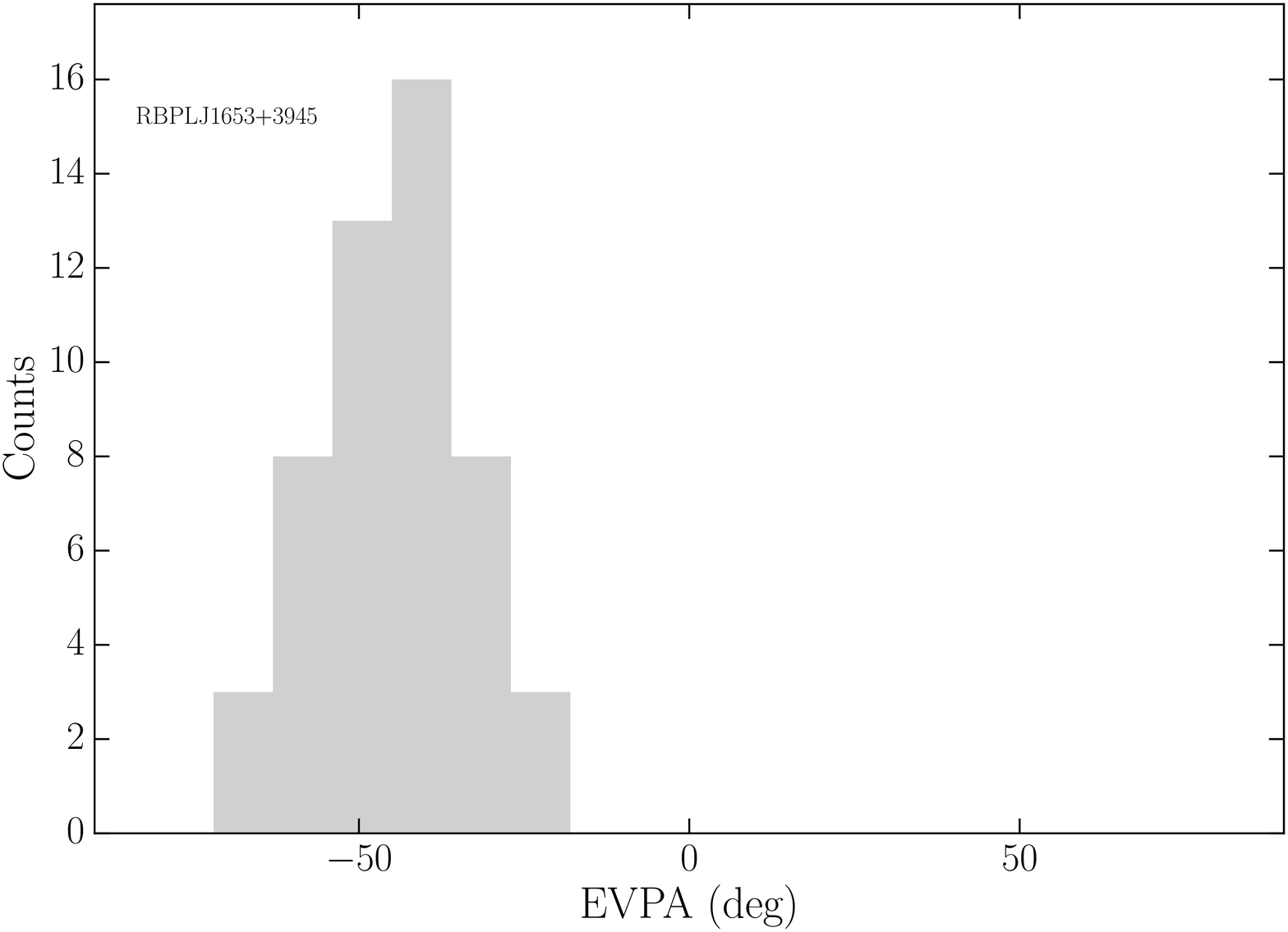}\\
  \includegraphics[trim=10pt 0pt 0pt 0pt  ,clip,width=0.4\textwidth]{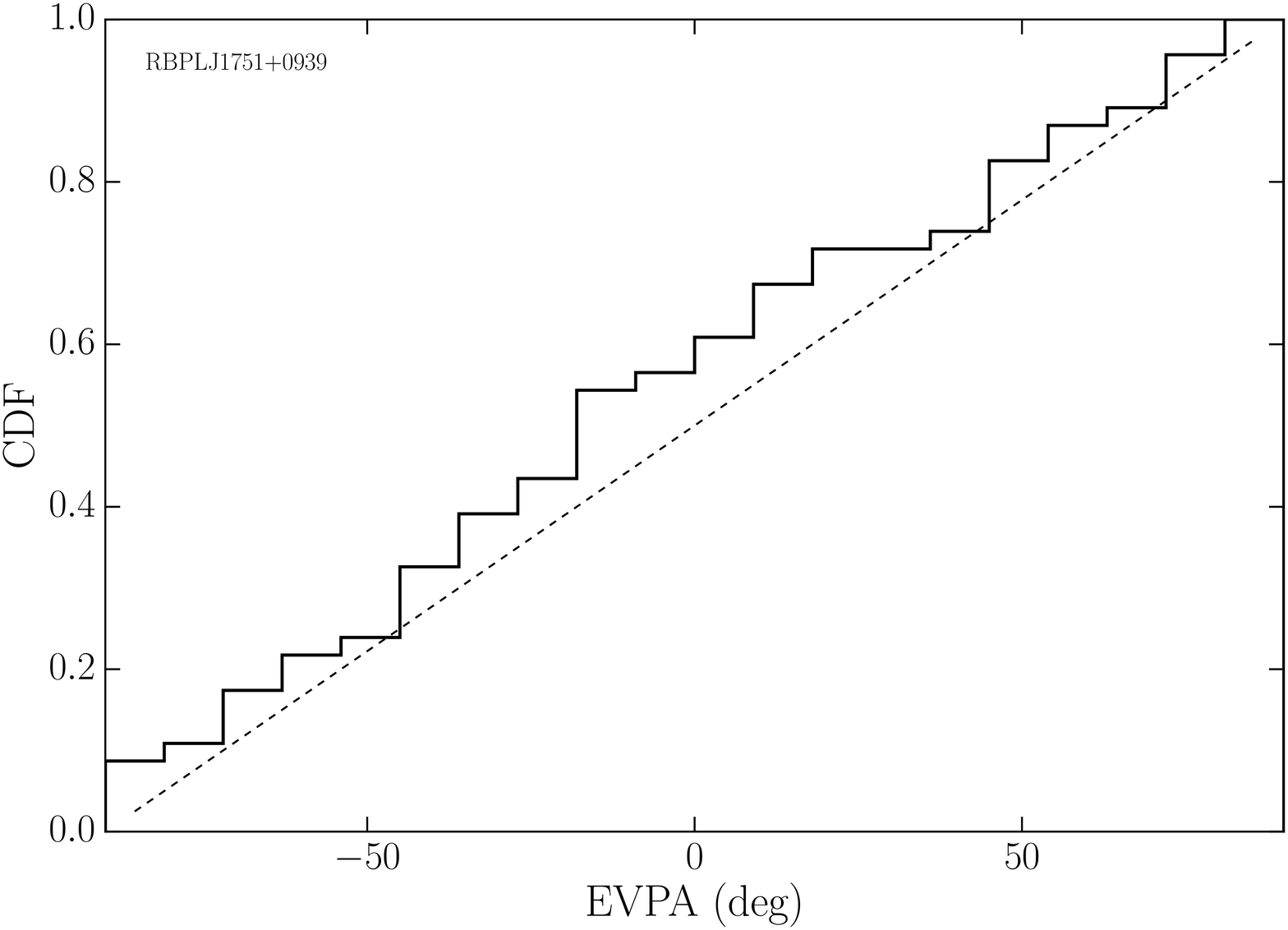}&\includegraphics[trim=10pt 0pt 0pt 0pt  ,clip,width=0.4\textwidth]{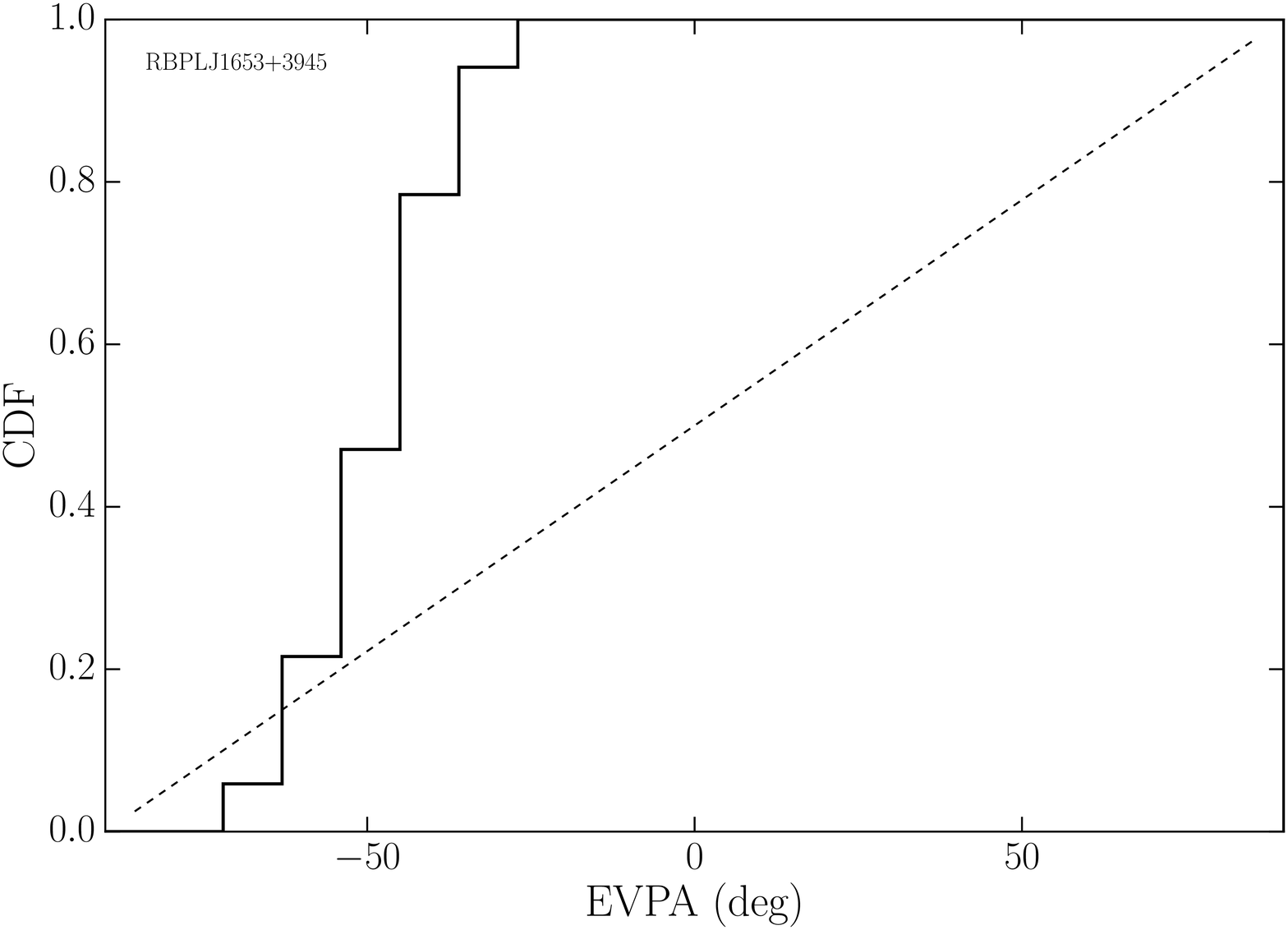}
\end{tabular}
\caption{
  The distribution of EVPA for a close-to-uniform case (highly random, RBPLJ1653$+$3945 {\it left}) and one
  case far-from-uniform (low randomness, RBPLJ1751$+$0939 {\it right}). {\it Upper row:} The distribution of
  EVPA. {\it Lower row:} The cumulative distribution function of the EVPA for those two cases (solid line) and
  the one of uniform distribution (dashed line). There are 46 data angle measurements for RBPLJ1751$+$0939 and
  51 for RBPLJ1653$+$3945.}
\label{fig:chi_peak_examples}
\end{figure*}
The polarization parameters have a strong dependence on the properties of the magnetic field
(e.g. uniformity). Given the relation between the polarization fraction and the synchrotron peak frequency
discussed above, we examine how the peak frequency may be influencing the behavior of the EVPA.

Figure~\ref{fig:chi_peak} demonstrates how well a uniform distribution describes the behavior of the EVPA of
each source as a function of the frequency of its synchrotron SED component peak. For every source we compute
the $\chi^2$ per degree of freedom, $\chi^2_\mathrm{red}$, between its angle distribution and a uniform
one. The computation has been done for 36 sources for which at least 20 measurements with
$\nicefrac{p}{\sigma_p}\ge 3$ are available so that a reliable estimate of the angle randomness can be
provided. Our calculations are done for 20 angle bins in the closed [$-90$,$+90$] interval. A large value of
$\chi^2_\mathrm{red}$ implies a big divergence from a uniform distribution and hence a low randomness of the
EVPA, which consequently centers around a preferred direction (e.g. Fig.~\ref{fig:chi_peak_examples}
right-hand column). The opposite is the case for small $\chi^2_\mathrm{red}$ values which imply a large
randomness of the EVPA that does not prefer any direction (e.g. Fig.~\ref{fig:chi_peak_examples} left-hand
column). The orange circles in Fig.~\ref{fig:chi_peak} mark the two exemplary cases shown in
Fig.~\ref{fig:chi_peak_examples}.

The Spearman's test does not support the presence of a monotonic relation between the EVPA randomness and the
synchrotron peak frequency ($\rho=0.34$, with a p-value $\sim0.044$). Two further tests, though, indicate a
dependency between the two parameters.

First, we classified our 36 sources as: low, intermediate and high-synchrotron peaked (LSP, ISP and HSP,
respectively). Then we selected 0.1 as the limiting value of $\chi^2_\mathrm{red}$ for a source to be
considered as non-uniform. We then found that: 11/14 (79\%) LSP, 7/14 (50\%) ISP and 3/8 (38\%) HSP sources,
have $\chi^2_\mathrm{red}$ below 0.1. Despite the small number statistics, this result indicates that HSP
sources are more likely to have a preferred and less variable EVPA than LSP sources.

Second, the {\it green} markers in Fig.~\ref{fig:chi_peak} show the mean $\chi^2_\mathrm{red}$ in each of five
synchrotron-peak frequency bins. The vertical error-bars show the spread of the values in the bin
($1\sigma$). A linear fit to the binned data -- the {\it green} dashed line -- gives a significant
slope of $0.037\pm0.010$.

We conclude that the randomness of the EVPA depends on the synchrotron peak frequency. The EVPA of HSP sources
is concentrated around preferred directions. The EVPA of LSP sources, on the other hand, is more variable and
less likely to have a preferred direction. In \S~\ref{sec:discussion} we argue that these two findings may be
evidence for a helical structure of the magnetic field.

\subsection{Polarization and source variability} 
\label{subsec:p_gamma_VI}

Depending on the mechanism producing the variability, it is likely that the degree of polarization relates to
the degree of variability at different bands. Here we examine the role that the radio and the optical
modulation indices may play.

\begin{figure}
\centering
\includegraphics[trim=10pt 0pt 0pt 0pt  ,clip,width=0.49\textwidth]{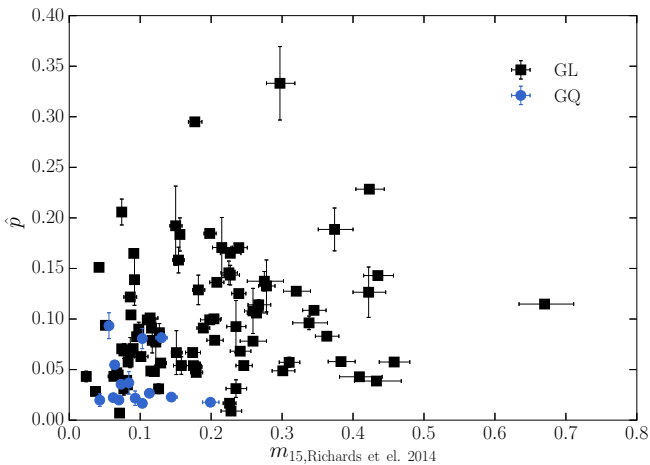} 
\caption{
  The median polarization fraction versus the 15~GHz intrinsic modulation index. In total we show 86 GL and 14
  GQ sources.}
\label{fig:mPD_mi}
\end{figure}

In Fig.~\ref{fig:mPD_mi} we plot the median polarization fraction versus the variability amplitude at 15~GHz
from \cite{2014MNRAS.438.3058R}, as that is quantified through the intrinsic modulation index introduced by
\cite{2011ApJS..194...29R}. As shown there, the two are correlated with Spearman's test giving a
$\rho\sim0.35$ and a p-value of about $3\times10^{-4}$. The GQ sources have preferentially low radio
modulation indices, as it was already found by \cite{2011ApJS..194...29R}. However, the GQ sources have
average polarization fractions that are low even compared to GL sources with comparable radio modulation
indices.

\begin{figure}
\centering
\includegraphics[trim=10pt 0pt 0pt 0pt  ,clip,width=0.49\textwidth]{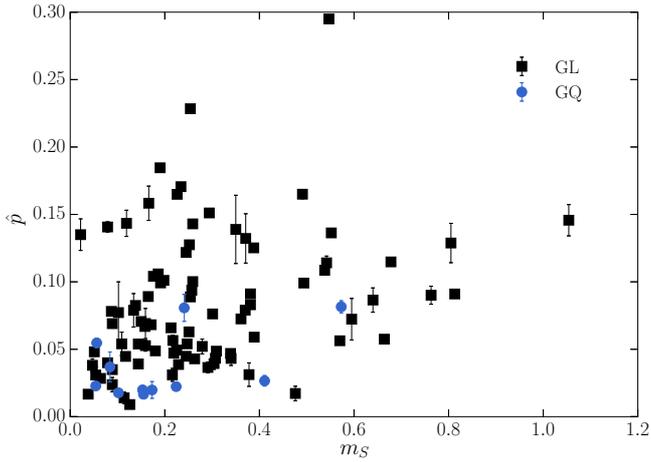} 
\includegraphics[trim=10pt 0pt 0pt 0pt  ,clip,width=0.49\textwidth]{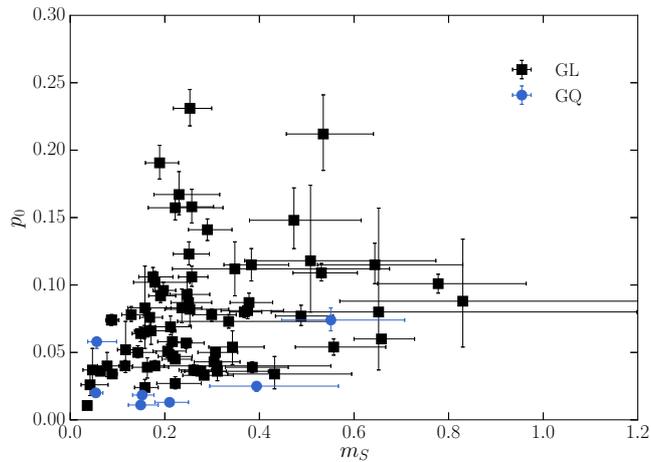} 
\caption{
The polarization fraction versus the $R$-band flux density modulation
  index. The upper panel is using the median polarization fraction $\hat{p}$ and the lower one the intrinsic
  mean $p_0$.}
\label{fig:mPD_miR}
\end{figure}

In Fig.~\ref{fig:mPD_miR} we examine the dependence of the polarization fraction on the variability amplitude
of the $R$-band flux density. In the upper panel we plot the observed median polarization fraction $\hat{p}$
and the $R$-band flux density modulation index $m_S$.  In this case Spearman's $\rho$, when including both GL
and GQ sources, is around $0.38$ with a p-value of $10^{-4}$, indicating a rather significant
correlation. Similarly, in the lower panel we show the maximum-likelihood intrinsic mean polarization fraction
$p_0$ and the $m_S$ which gave a Spearman's $\rho\approx 0.38$ with a p-value of $8\times10^{-4}$. Again, GQ
sources are systematically less polarized on average than sources with comparable optical modulation indices.

\begin{figure}
\centering
\includegraphics[trim=10pt 0pt 0pt 0pt  ,clip,width=0.49\textwidth]{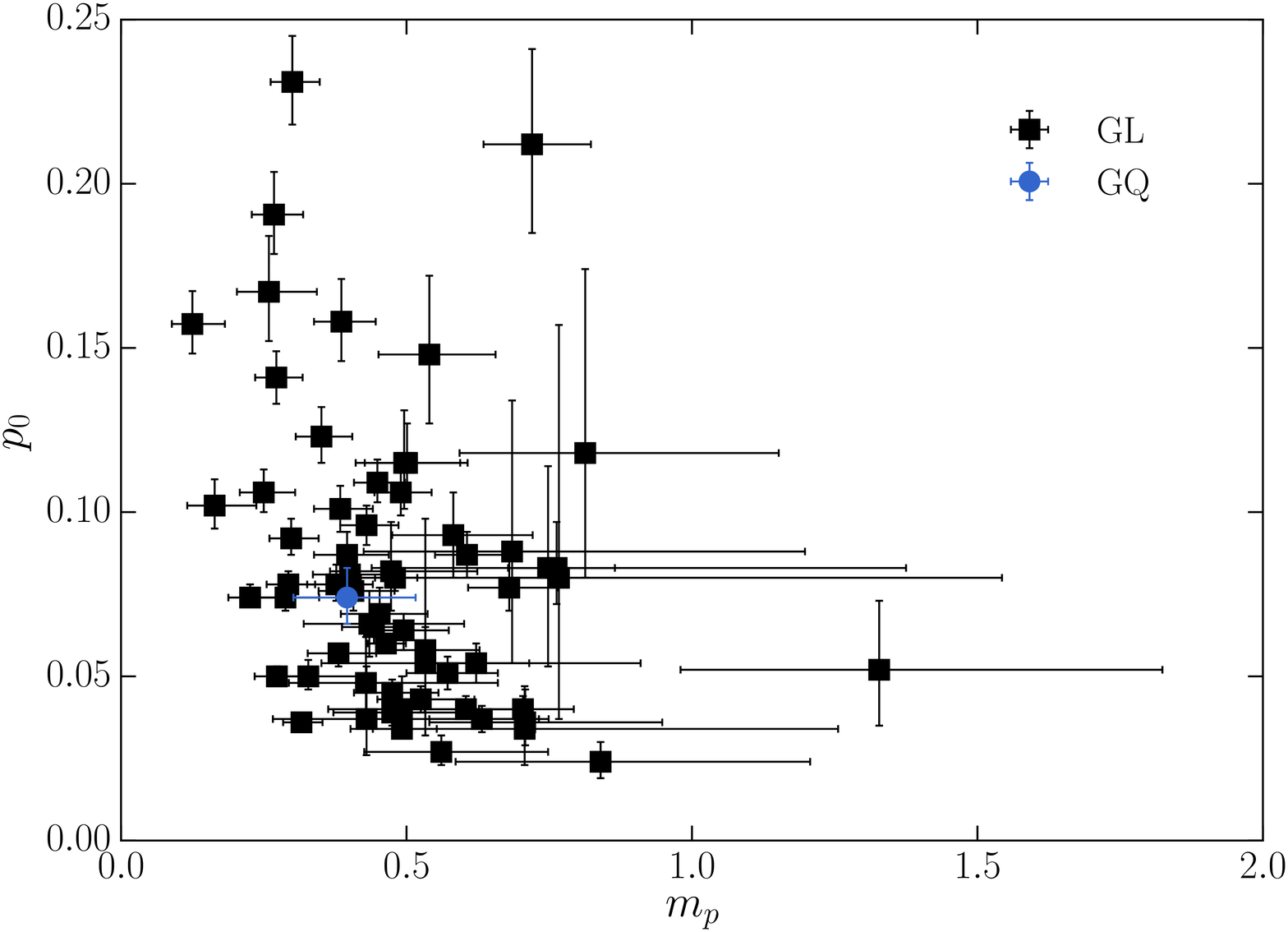} 
\caption{
The mean intrinsic polarization fraction versus the intrinsic polarization
  modulation index.}
\label{fig:p0_mp}
\end{figure}

Finally, in Fig.~\ref{fig:p0_mp} we examine whether $p_0$ depends on the amplitude of the variability
quantified through the intrinsic polarization modulation index $m_p$. Spearman's test gave a $\rho$ of around
$-0.31$ with a significance of p-value $0.013$.

We conclude that the variability amplitude, in both radio and optical flux density, affects the mean observed
polarization. With comparable Spearman's test results, higher polarization is associated with stronger
variability in either the optical or the radio. Finally, there is also a weak indication that stronger
variability in optical polarization associates (on average) with lower polarization although of lower
significance. Nevertheless, these correlations cannot explain GL-GQ polarization dichotomy.

\subsection{The polarization variability of the GL and GQ samples} 
\label{sec:pol_var}
Intrigued by the dichotomy between GL and GQ samples in terms of their polarization fraction and given the
correlation between the $\hat{p}$ and the $R$-band modulation index (Fig.~\ref{fig:mPD_miR}), we have searched
for a similar dichotomy in the distribution of their polarization variability. We also consider its dependence
of redshift.

\begin{figure}
\centering
\includegraphics[trim=10pt 0pt 0pt 0pt  ,clip,width=0.49\textwidth]{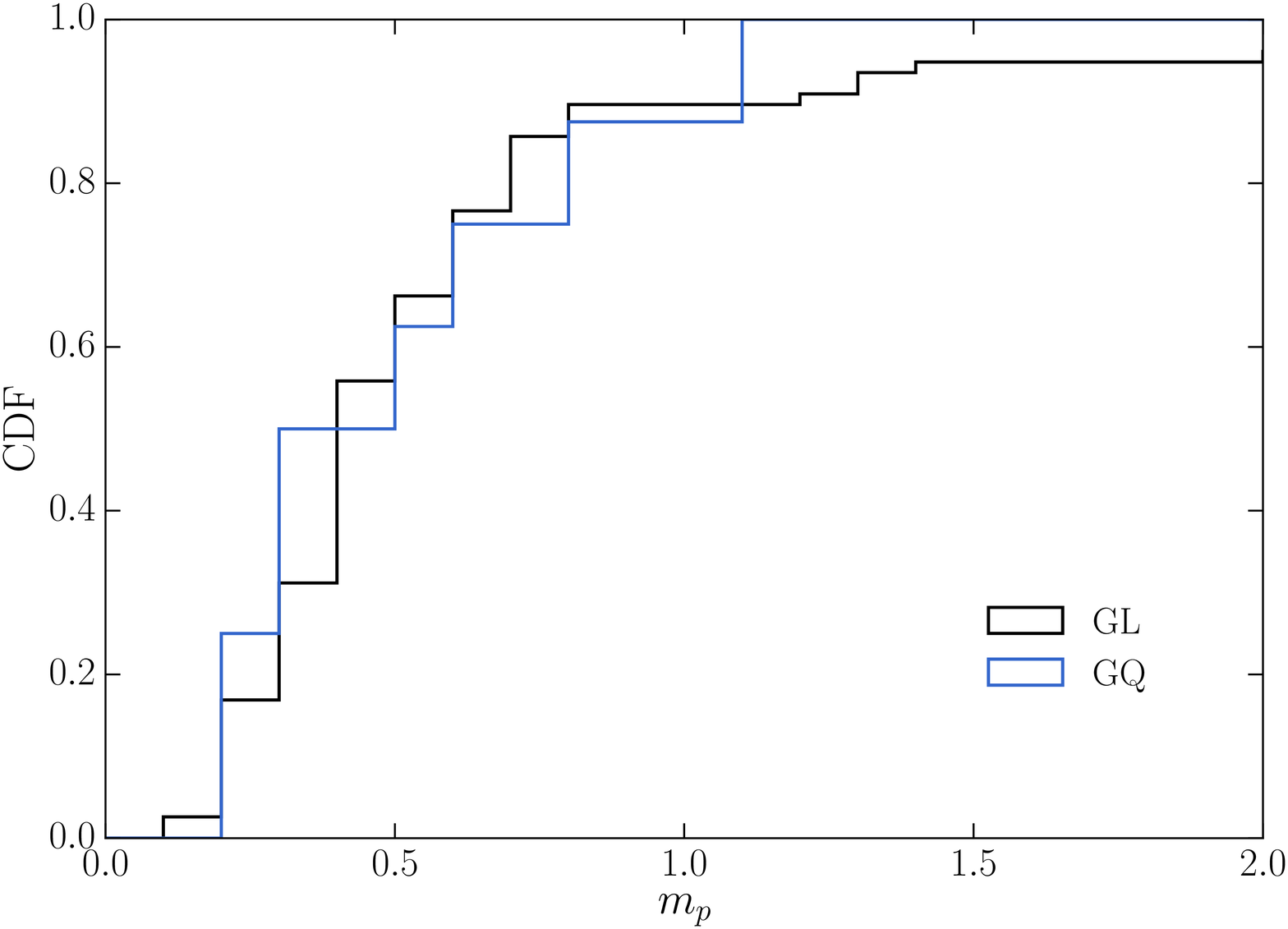} 
\caption{
  The intrinsic modulation index $m_p$ for the GL and GQ samples. In cases where the $m_p$ was not available
  $2\sigma$ upper limits have been included instead.}
\label{fig:mp-distr.eps}
\end{figure}

The distribution of the intrinsic modulation index $m_p$ is shown in Fig.~\ref{fig:mp-distr.eps} (77 GL and 8
GQ sources). Of these, 13 GL and 6 GQ sources have only $2\sigma$ upper limits were available. A standard
two-sample K-S test could not distinguish the two distributions ($D=0.36$ and p-value of $0.255$). A Gehan's
generalized Wilcoxon test indicated a similar result with a p-value of $0.167$.

\begin{figure}
\centering
\includegraphics[trim=10pt 0pt 0pt 0pt  ,clip,width=0.49\textwidth]{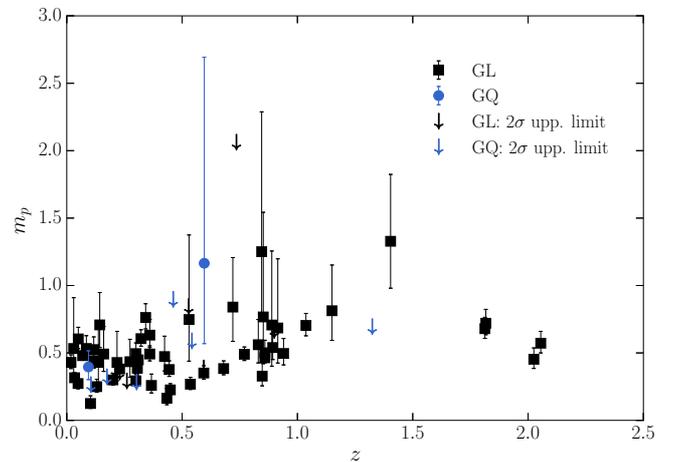} 
\caption{
  The intrinsic modulation index of the polarization fraction versus the redshift. The arrows indicate
  $2\sigma$ upper limits. The $y$-axis has been truncated at 3 excluding three GL upper limits close to 3.5, 4
  and 7.}
\label{fig:mp-z.eps}
\end{figure}

Contrary to the median polarization fraction $\hat{p}$ (see Sect.~\ref{subsec:p-z}), the intrinsic modulation
index $m_p$ depends on the source redshift. In Fig.~\ref{fig:mp-z.eps} the arrows indicate $2\sigma$ upper
limits. A Spearman test for collectively the GL and the GQ sources, excluding the upper limits, gave a $\rho$
of 0.43 and a p-value of $10^{-3}$. When the upper limits are included (10 GL and 6 GQ sources) the
correlation remains as tight ($\rho\approx 0.42$) but the significance improves by almost one order of
magnitude with a p-value of $2\times 10^{-4}$.

We conclude that although there is no dichotomy between the polarization variability index $m_p$ of GL and GQ
sources similar to the one seen for $\hat{p}$, a significant correlation exists between $m_p$ and redshift.

\subsection{Variability of optical flux density and polarization against the variability in other bands
 } 
\label{subsect:mp_mi_VI}
We are now interested in examining whether the variability in the $R$-band, both in total flux density and in
fractional polarization, correlates with the variability in other bands. That would be expected in the radio
and the optical if photons in those bands belong to the same synchrotron component.

\begin{figure}
\centering
\includegraphics[trim=10pt 0pt 0pt 0pt  ,clip,width=0.49\textwidth]{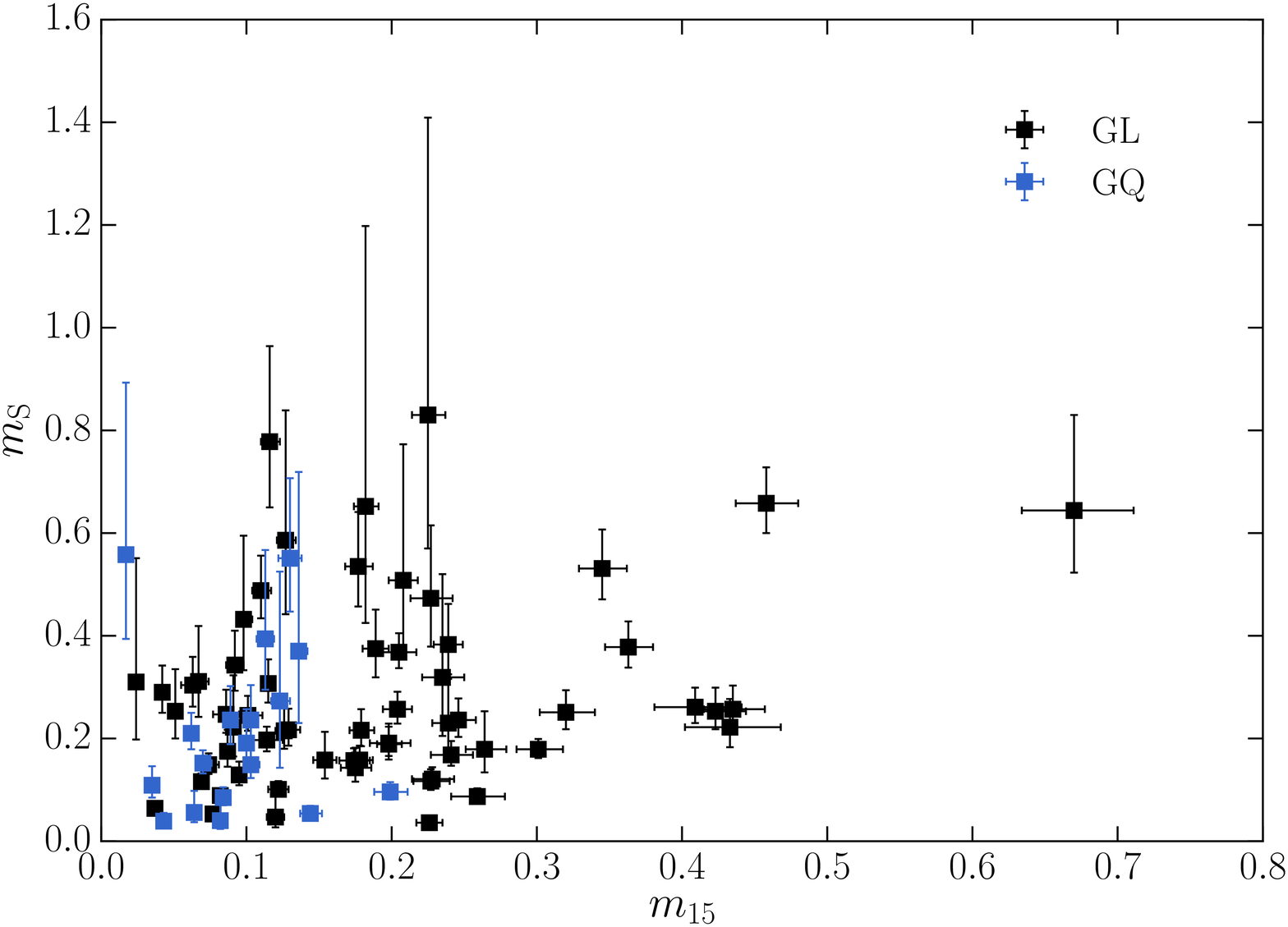} 
\caption{
The $R$-band intrinsic modulation index $m_S$ versus that at 15 GHz, $m_\mathrm{15}$. }
\label{fig:miS_m15}
\end{figure}

For a total of 61 GL and 18 GQ sources estimates for both $m_{15}$ \citep{2014MNRAS.438.3058R} and $m_S$ are
available. Those are shown in Fig.~\ref{fig:miS_m15} and as it appears they are not correlated (Spearman's
$\rho\approx 0.25$ with p-value $0.025$).

\begin{figure}
\centering
\includegraphics[trim=10pt 0pt 0pt 0pt  ,clip,width=0.49\textwidth]{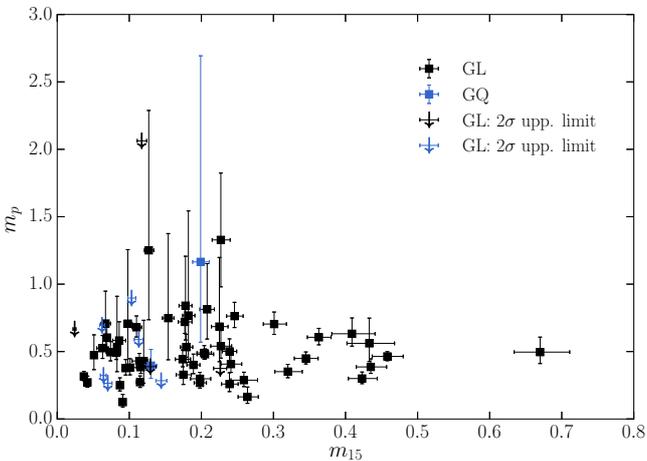} 
\caption{
The intrinsic polarization modulation index $m_p$ versus that at 15 GHz, $m_\mathrm{15}$. The arrows mark $2\sigma$ upper
  limits. The $y$-axis is truncated at 3 excluding one upper limit around 7.}
\label{fig:miPD_mi}
\end{figure}

Figure~\ref{fig:miPD_mi} shows the intrinsic polarization modulation index $m_p$ versus $m_{15}$. Whenever
possible, $2\sigma$ upper limits are also shown. As in the $m_{15}$--$m_S$ case, there is no clear
correlation, implying that the amplitude of the 15 GHz total intensity variability, is not connected to the
variability amplitude of the optical polarization fraction. The validity of this conclusion, of course, relies
on the assumption that the radio and optical data sets used carry the characteristics of the variability
mechanisms even though they are not contemporaneous.

\begin{figure}
\centering
\includegraphics[trim=10pt 0pt 0pt 0pt  ,clip,width=0.49\textwidth]{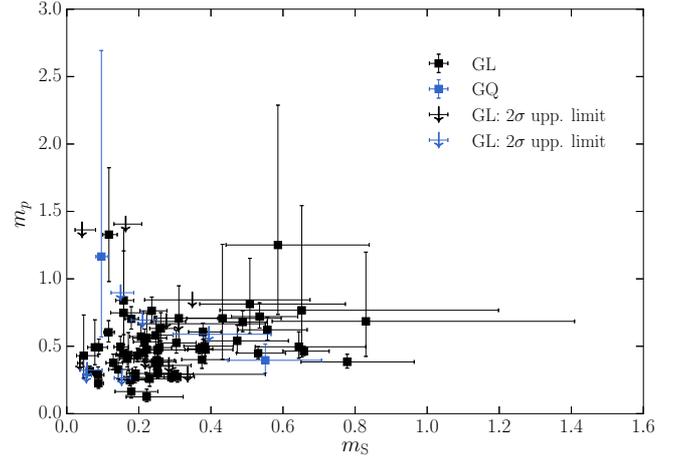} 
\caption{
The median polarization fraction versus the modulation index of the $R$-band flux
  density, $m_S$. The $y$-axis has been truncated at 3 excluding 3 upper limits at around 3.5, 4 and 7. }
\label{fig:mPD_Rmi}
\end{figure}

There is a weak indication of a possible mild correlation between the intrinsic polarization variability index
$m_p$ and the flux density variability index $m_\mathrm{S}$ Fig.~\ref{fig:mPD_Rmi}. Using the GL sources alone
gave a $\rho$ around 0.3 although with a significance below the $2.5\sigma$ level (p-value$\approx0.016$).

\section{SUMMARY AND DISCUSSION} 
\label{sec:discussion}

We have presented the average polarimetric and photometric properties and the variability parameters, of GL
and GQ sources observed with \rbpl~during the first two observing seasons. Our analysis concentrated on (a)
quantifying the possible difference in the polarization of the GL and GQ sources that was first found by
\cite{2014MNRAS.442.1693P}; and (b) investigating its possible causes. We also examined whether the
polarization variability shows a similar dichotomy for GL and GQ sources. We have found that:

\paragraph*{{\it The average polarization does not depend on luminosity. }}
While in the Survey Paper the un-polarized starlight contribution of the host galaxy was suggested as being
possibly responsible for the apparent de-polarization of the brightest sources, a more detailed analysis in
luminosity space revealed that sources that are both very luminous and highly polarized are possible (see
Fig.~\ref{fig:PDvsL_k}).

\paragraph*{{\it The average polarization fraction of GL and GQ sources differs. }} The two samples have
different mean polarization fractions: the distributions of $\hat{p}$ are different at an almost $4\sigma$
level, while those of the intrinsic mean polarization fraction $p_0$ have yielded a significance of
$\sim3\sigma$. A Gehan's generalized Wilcoxon test applied on a dataset including $2\sigma$ upper limits in
$p_0$, produces a similar result (Fig.~\ref{fig:PD-GLoudns} lower panel). A log-normal distribution
fit to the two distributions of $p_0$ gives the mean intrinsic polarization $\left<p_0\right>$ of
$(9.2\pm0.8)\times 10^{-2}$ for GL and $(3.1\pm0.8)\times10^{-2}$ for GQ sources.

\paragraph*{{\it The variability amplitude of the polarization fraction does not differ between GL and GQ sources. }}
Unlike the polarization fraction, its variability amplitude does not show the same dichotomy between GL and GQ
samples. However the sample consisted of 64 GL and 2 GQ sources (of which 19 have only upper limits), so small
number statistics may limit our ability to establish a difference between the two populations.  This makes any
conclusion concerning the distributions of $m_p$ ambiguous. However, the very fact that for the majority of GQ
sources we were able only to place upper limits on the amplitude of optical polarization variability may be
seen as an indication that GQ sources are less variable. That is indeed the case in terms of radio and optical
flux density modulation index as Fig.~\ref{fig:mPD_mi}~and~\ref{fig:mPD_miR}, show.

\paragraph*{{\it The stronger the variability in radio or optical the larger the mean polarization. }}
Figures~\ref{fig:mPD_mi}~and~\ref{fig:mPD_miR} suggest that the larger the amplitude of the radio and the
$R$-band flux density variability, the higher is the median polarization. On the other hand, the polarization
variability amplitude $m_p$ does not seem to influence the median polarization although there is even an
indication that the two are anti-correlated (Fig.~\ref{fig:p0_mp}). We have also examined whether the high
energy (2FGL) variability index is influencing the polarization fraction and found no evidence for such a
dependence.

\paragraph*{{\it The modulation index of the polarization fraction is redshift dependent. }}
Contrary to the polarization fraction itself, its variability amplitude seems to be a function of redshift.

\paragraph*{{\it  Source class is not the reason for the GL-GQ dichotomy.}}
The dominance of radio quasars in the GQ sample could explain the observed dichotomy, if BL~Lac objects and
Flat Spectrum Radio Quasars were characterized by different distributions of $p$. A two-sample K-S test
between quasars and BL~Lac objects has shown that the two distributions are indistinguishable.  It must be noted
however that the GQ sources reach larger redshifts (Fig.~\ref{fig:PDF-z}) which could potentially have an
effect on the gamma-ray detectability given the maximum redshift that {\it Fermi} can probe. Our findings
however cannot be influenced by this; (a) because GQ sources for which $\hat{p}$ values are available and
hence are included in our plots, are limited to $z< 1.5$; and (b) as can be seen in Fig.~\ref{fig:mPD_z.eps},
the degree of polarization is independent of the source cosmological distance.

\paragraph*{{\it The optical polarization fraction and the randomness of the polarization angle, depend on the synchrotron peak frequency.}}
Figure~\ref{fig:PD_peak} revealed a synchrotron-peak-dependent envelope limiting the polarization fraction:
the fractional polarization $\hat{p}$ of LSP sources is on average higher {than that} for HSP ones,
while their polarization spreads over a broader range extending to considerably higher values of $\hat{p}$. We
{have shown} that if we exclude the GQ sources (for which the synchrotron peak is severely
under-sampled), there is a significant anti-correlation between $\hat{p}$ and the rest-frame frequency of the
synchrotron peak, $\nu_\mathrm{s}$.  The anti-correlation {becomes} clearer and more significant when
only the ``bzb'' subset of the GL sample {is} considered.  A similar relation between the fractional
polarization of the VLBA core and the synchrotron peak frequency {has been found} by
\cite{2011ApJ...742...27L}. When they { have focused} only on LSP and HSP BL~Lac objects that span
similar redshift ranges, they observe the same trend. They explain the observed correlation as a result of the
balance between the intrinsic gamma-ray loudness and the Doppler boosting of the sources given the general
association of high polarization to highly Doppler-boosted jets. Myserlis et al. (in prep.) look at the
fractional polarization of roughly 35 {\it Fermi} sources and find that at 2.64 and 4.85~GHz the same relation
is apparent. Specifically at 4.85~GHz they find that Spearman's $\rho=-0.35$.

We also {show} that apart from the polarization fraction, the randomness of the EVPA depends on the
synchrotron peak frequency. LSP sources tend to show a random orientation of their, unlike HSP sources which
tend to show a preferred direction.

\subsection{A qualitative interpretation of the observed trends}
\begin{figure}
\centering
\includegraphics[trim=0pt 0pt 0pt 0pt  ,clip,width=0.38\textwidth]{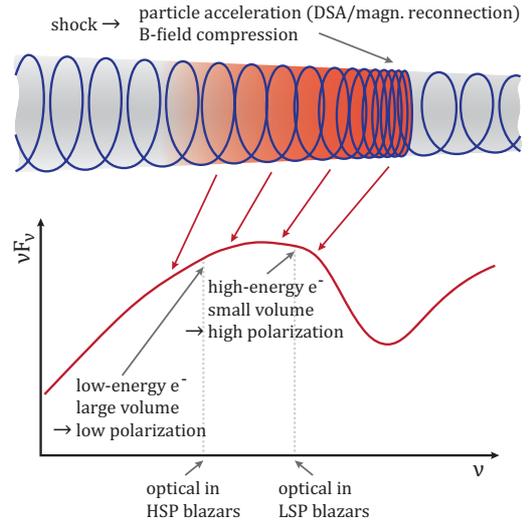} 
\caption{Cartoon representation of the shock-in-jet scenario. The downstream directions is towards the left.}
\label{fig:shock}
\end{figure}

In this section, we propose a simple, qualitative explanation for the various trends of the average degree of
polarization found in this study. It is based on a basic shock-in-jet scenario, as sketched in
Fig.~\ref{fig:shock}. The jet is expected to be pervaded by a helical magnetic field structure, on which a
turbulent $B$-field component is superposed. A mildly relativistic shock, caused either by a static
disturbance in the environment of the jet (i.e. a standing shock), or by the collision of plasmoids
propagating along the jet with different Lorentz factors (internal shock), mediates efficient particle
acceleration due to diffusive shock acceleration (DSA) or magnetic reconnection in a small volume,
concentrated in the immediate downstream environment of the shock. As particles are advected away from the
shock, they cool, primarily due to the emission of synchrotron and Compton radiation. Consequently, the
highest-energy particles, responsible for the emission near and beyond the peak of the synchrotron (and
Compton) SED components, are expected to be concentrated in a small volume immediately downstream of the
shock, where the shock-compressed magnetic field is expected to have a strong ordered (helical) component, in
addition to shock-generated turbulent magnetic fields. Substantial degrees of polarization are thus expected
near and beyond the peak of the synchrotron SED component. Due to progressive cooling of shock-accelerated
electrons as they are advected downstream, the volume from which lower-frequency synchrotron emission is
received, is expected to increase monotonically with decreasing frequency. One therefore expects a lower
degree of polarization with decreasing frequency due to de-polarization from the superposition of radiation
zones with different $B$-field orientations.

First of all, the general trend of a higher degree of polarization for GL compared to GQ AGN, may be explained
as follows: GL AGN (i.e., primarily blazars) are known to be highly variable, indicating a strong jet
dominance throughout most of the SED due to a high degree of Doppler boosting
\citep[e.g.][]{2010A&A...512A..24S,2015ApJ...810L...9L} and the frequent occurrence of impulsive particle
acceleration events, such as the shock-in-jet scenario described above.  On the other hand, GQ AGN appear to
represent objects in which Doppler boosting is less extreme and/or impulsive particle acceleration episodes
are less efficient, thus not accelerating particles to the energies required for gamma-ray production at
measurable levels. Consequently, optical synchrotron emission is likely to be produced on larger volumes than
in the more active GL objects, thus naturally explaining the lower degree of polarization.

This scenario also naturally explains the dependence of the degree of polarization on the synchrotron peak
frequency: In LSP blazars, such as FSRQs and low-frequency peaked BL~Lacs (LBLs), the synchrotron peak
frequency is typically located in the infrared.  Thus, the optical regime represents the high-frequency
portion of the synchrotron emission, for which -- as elaborated above -- one expects a high degree of
polarization.  On the contrary, in HSP blazars, such as high-frequency peaked BL~Lacs (HBLs), the synchrotron
peak tends to be located at UV or X-ray frequencies. Thus, here the optical regime represents the
low-frequency part of the synchrotron SED, for which one expects a lower degree of polarization.

Finally, this scenario also explains the tendency of the optical EVPA rotation events to occur preferentially
in LSP sources as we present elsewhere (Blinov et al. in prep.). In the case of LSP sources the optical
emission originates at the small volume in the immediately downstream environment of the shock, where the
magnetic field has a strong helical component. In HSP sources on the other hand, the optical emission
originates in a larger region farther downstream of the shock, where the electrons have already lost part of
their energy and the turbulent $B$-field component becomes more significant. It has been shown by
\cite{2015MNRAS.453.1669B} and \cite{2016arXiv160300249K} that two types of EVPA rotations may coexist in
blazars. The smooth deterministic EVPA rotations may occur preferentially when plasmoids propagate through
regions where the helical field component is dominant \citep[e.g.][]{2008Natur.452..966M,2010ApJ...710L.126M,
  2014ApJ...789...66Z, 2015ApJ...804...58Z}, {whereas} further downstream the EVPA variability is more likely to
be driven by stochastic processes. Consequently, smooth rotations are more likely to occur in LSP than HSP
sources. Indeed, all five rotations in Fig. 8 of \cite{2015MNRAS.453.1669B} associated with strong gamma-ray
flares and short time lag from the flare, which are hence considered deterministic, have occurred in LSP
sources. Moreover, the optical emission region in LSP sources is smaller than in HSP sources and thus expected
to be more variable. In the context of stochastic variations, larger emitting region implies an increased
number of cells, which decreases the variability \citep[e.g.][]{2016arXiv160300249K}. Also, the larger
emission region in HSP sources increases the variability time scale.

Assuming the superposition of a helical magnetic field component and a turbulent one, LSP and HSP sources may
have an underlying, stable EVPA component due to the helical field component. In LSP sources the stable
component may not be clearly visible owing to stronger variability and shorter variability time scales. In HSP
sources, in which the variability amplitudes are lower and variability time scales are longer, the stable
component may be more dominant. There, the combination of local turbulence that keeps the global magnetic
field structure intact can explain a preferred, though slightly variable EVPA. Only long term observations can
confirm whether the EVPA has a truly preferred orientation on time scales longer than the \rbpl~observing
periods.

If the difference between LSP and HSP sources in terms of polarization is indeed caused merely by the fact
that observations in the optical band probe (a) regions of different size and (b) different parts of the
particle distribution, then we would expect the same polarization variability in HSP sources at X-ray bands as
in LSP sources in optical bands.

It is worth noting that in this scenario the rotations of the EVPA are expected to be happening downstream the
shock in contrast to earlier suggestions \citep[e.g.][]{2010ApJ...710L.126M} that the region responsible for
these events was just upstream of the shock.

\section*{Acknowledgements}

The \rbpl~project is a collaboration between Caltech in the USA, MPIfR in Germany, Toru\'{n} Centre for
Astronomy in Poland, the University of Crete/FORTH in Greece, and IUCAA in India. The U. of Crete group
acknowledges support by the ``RoboPol'' project, which is implemented under the ``Aristeia'' action of the
``Operational Programme Education and Lifelong Learning'' and is co-funded by the European Social Fund (ESF),
Greek National Resources, the European Commission Seventh Framework Program (FP7) through grants
PCIG10-GA-2011-304001 ``JetPop'' and PIRSES-GA-2012-31578 ``EuroCal''. This research was also supported in
part by NASA grant NNX11A043G and NSF grant AST-1109911, the Polish National Science Centre, grant number
2011/01/B/ST9/04618 and the COST Action MP1104 "Polarization as a tool to study the Solar System and
beyond". KT acknowledges support by the European Commission Seventh Framework Program (FP7) through the Marie
Curie Career Integration Grant PCIG-GA-2011-293531 ``SFOnset''. IM and SK were funded by the International Max
Planck Research School (IMPRS) for Astronomy and Astrophysics at the Universities of Bonn and
Cologne. M.~Balokovi\'{c} acknowledges support from the International Fulbright Science and Technology
Award. TH was supported in part by the Academy of Finland project number 267324. We would also like to
acknowledge partial support from the EU FP7 Grant PIRSES- GA-2012-316788. The work of M. Boettcher is
supported by the South African Research Chair Initiative of the Department of Science and Technology and the
National Research Foundation\footnote{Any opinion, finding and conclusion or recommendation expressed in this
  material is that of the authors and the NRF does not accept any liability in this regard.}  of South Africa.
This research has made use of the NASA/IPAC Extragalactic Database (NED) which is operated by the Jet
Propulsion Laboratory, California Institute of Technology, under contract with the National Aeronautics and
Space Administration. It also made use of Astropy, \url{http://www.astropy.org}, a community-developed core
Python package for Astronomy \citep{astropy}. EA wishes to thank Dr. R. Porcas and Dr. F. Mantovani of the
MPIfR for the fruitful discussions on the statistical behavior of the polarization parameters, and the MPIfR
internal referee Dr. B. Boccardi for the constructive comments on the manuscript. Finally, EA wants to thank
the anonymous referee for the careful reading and numerous insightful comments of the first manuscript.




\bibliographystyle{mnras}
\bibliography{References} 

\begin{thebibliography}{}
\makeatletter
\relax
\def\mn@urlcharsother{\let\do\@makeother \do\$\do\&\do\#\do\^\do\_\do\%\do\~}
\def\mn@doi{\begingroup\mn@urlcharsother \@ifnextchar [ {\mn@doi@}
  {\mn@doi@[]}}
\def\mn@doi@[#1]#2{\def\@tempa{#1}\ifx\@tempa\@empty \href
  {http://dx.doi.org/#2} {doi:#2}\else \href {http://dx.doi.org/#2} {#1}\fi
  \endgroup}
\def\mn@eprint#1#2{\mn@eprint@#1:#2::\@nil}
\def\mn@eprint@arXiv#1{\href {http://arxiv.org/abs/#1} {{\tt arXiv:#1}}}
\def\mn@eprint@dblp#1{\href {http://dblp.uni-trier.de/rec/bibtex/#1.xml}
  {dblp:#1}}
\def\mn@eprint@#1:#2:#3:#4\@nil{\def\@tempa {#1}\def\@tempb {#2}\def\@tempc
  {#3}\ifx \@tempc \@empty \let \@tempc \@tempb \let \@tempb \@tempa \fi \ifx
  \@tempb \@empty \def\@tempb {arXiv}\fi \@ifundefined
  {mn@eprint@\@tempb}{\@tempb:\@tempc}{\expandafter \expandafter \csname
  mn@eprint@\@tempb\endcsname \expandafter{\@tempc}}}

\bibitem[\protect\citeauthoryear{{Abdo} et~al.,}{{Abdo}
  et~al.}{2010}]{2010Natur.463..919A}
{Abdo} A.~A.,  et~al., 2010, \mn@doi [\nat] {10.1038/nature08841}, \href
  {http://adsabs.harvard.edu/abs/2010Natur.463..919A} {463, 919}

\bibitem[\protect\citeauthoryear{{Acero} et~al.,}{{Acero}
  et~al.}{2015}]{2015ApJS..218...23A}
{Acero} F.,  et~al., 2015, \mn@doi [\apjs] {10.1088/0067-0049/218/2/23}, \href
  {http://adsabs.harvard.edu/abs/2015ApJS..218...23A} {218, 23}

\bibitem[\protect\citeauthoryear{{Ackermann} et~al.,}{{Ackermann}
  et~al.}{2015}]{2015ApJ...810...14A}
{Ackermann} M.,  et~al., 2015, \mn@doi [\apj] {10.1088/0004-637X/810/1/14},
  \href {http://adsabs.harvard.edu/abs/2015ApJ...810...14A} {810, 14}

\bibitem[\protect\citeauthoryear{{Ahn} et~al.,}{{Ahn} et~al.}{2012}]{ahn12}
{Ahn} C.~P.,  et~al., 2012, \mn@doi [\apjs] {10.1088/0067-0049/203/2/21}, \href
  {http://adsabs.harvard.edu/abs/2012ApJS..203...21A} {203, 21}

\bibitem[\protect\citeauthoryear{{Aleksi{\'c}} et~al.,}{{Aleksi{\'c}}
  et~al.}{2014}]{2014A&A...567A.135A}
{Aleksi{\'c}} J.,  et~al., 2014, \mn@doi [\aap] {10.1051/0004-6361/201423364},
  \href {http://cdsads.u-strasbg.fr/abs/2014A%26A...567A.135A} {567, A135}

\bibitem[\protect\citeauthoryear{{Andersson}, {Lazarian}  \&
  {Vaillancourt}}{{Andersson} et~al.}{2015}]{2015ARA&A..53..501A}
{Andersson} B.-G.,  {Lazarian} A.,   {Vaillancourt} J.~E.,  2015, \mn@doi
  [\araa] {10.1146/annurev-astro-082214-122414}, \href
  {http://adsabs.harvard.edu/abs/2015ARA%26A..53..501A} {53, 501}

\bibitem[\protect\citeauthoryear{{Andruchow}, {Cellone}  \&
  {Romero}}{{Andruchow} et~al.}{2008}]{2008MNRAS.388.1766A}
{Andruchow} I.,  {Cellone} S.~A.,   {Romero} G.~E.,  2008, \mn@doi [\mnras]
  {10.1111/j.1365-2966.2008.13502.x}, \href
  {http://adsabs.harvard.edu/abs/2008MNRAS.388.1766A} {388, 1766}

\bibitem[\protect\citeauthoryear{{Astropy Collaboration}}{{Astropy
  Collaboration}}{2013}]{astropy}
{Astropy Collaboration} 2013, arXiv.org, p. \mbox{1307.6212}

\bibitem[\protect\citeauthoryear{{Blandford} \& {K{\"o}nigl}}{{Blandford} \&
  {K{\"o}nigl}}{1979}]{1979ApJ...232...34B}
{Blandford} R.~D.,  {K{\"o}nigl} A.,  1979, \mn@doi [\apj] {10.1086/157262},
  \href {http://adsabs.harvard.edu/abs/1979ApJ...232...34B} {232, 34}

\bibitem[\protect\citeauthoryear{{Blinov} et~al.,}{{Blinov}
  et~al.}{2015}]{2015MNRAS.453.1669B}
{Blinov} D.,  et~al., 2015, \mn@doi [\mnras] {10.1093/mnras/stv1723}, \href
  {http://adsabs.harvard.edu/abs/2015MNRAS.453.1669B} {453, 1669}

\bibitem[\protect\citeauthoryear{{Blinov} et~al.,}{{Blinov}
  et~al.}{2016}]{2016MNRAS.457.2252B}
{Blinov} D.,  et~al., 2016, \mn@doi [\mnras] {10.1093/mnras/stw158}, \href
  {http://adsabs.harvard.edu/abs/2016MNRAS.457.2252B} {457, 2252}

\bibitem[\protect\citeauthoryear{{Burrows} et~al.,}{{Burrows}
  et~al.}{2005}]{burrows05}
{Burrows} D.~N.,  et~al., 2005, \mn@doi [\ssr] {10.1007/s11214-005-5097-2},
  \href {http://adsabs.harvard.edu/abs/2005SSRv..120..165B} {120, 165}

\bibitem[\protect\citeauthoryear{{Condon}, {Cotton}, {Greisen}, {Yin},
  {Perley}, {Taylor}  \& {Broderick}}{{Condon} et~al.}{1998}]{Condon1998AJ}
{Condon} J.~J.,  {Cotton} W.~D.,  {Greisen} E.~W.,  {Yin} Q.~F.,  {Perley}
  R.~A.,  {Taylor} G.~B.,   {Broderick} J.~J.,  1998, \mn@doi [\aj]
  {10.1086/300337}, \href
  {http://adsabs.harvard.edu/cgi-bin/nph-bib_query?bibcode=1998AJ....115.1693C&db_key=AST}
  {115, 1693}

\bibitem[\protect\citeauthoryear{{Dunlop}, {McLure}, {Kukula}, {Baum}, {O'Dea}
  \& {Hughes}}{{Dunlop} et~al.}{2003}]{Dunlop03}
{Dunlop} J.~S.,  {McLure} R.~J.,  {Kukula} M.~J.,  {Baum} S.~A.,  {O'Dea}
  C.~P.,   {Hughes} D.~H.,  2003, \mn@doi [\mnras]
  {10.1046/j.1365-8711.2003.06333.x}, \href
  {http://adsabs.harvard.edu/abs/2003MNRAS.340.1095D} {340, 1095}

\bibitem[\protect\citeauthoryear{{Fiorucci}, {Ciprini}  \& {Tosti}}{{Fiorucci}
  et~al.}{2004}]{2004A&A...419...25F}
{Fiorucci} M.,  {Ciprini} S.,   {Tosti} G.,  2004, \mn@doi [\aap]
  {10.1051/0004-6361:20034218}, \href
  {http://adsabs.harvard.edu/abs/2004A%26A...419...25F} {419, 25}

\bibitem[\protect\citeauthoryear{{Fukugita}, {Shimasaku}  \&
  {Ichikawa}}{{Fukugita} et~al.}{1995}]{fukugita95}
{Fukugita} M.,  {Shimasaku} K.,   {Ichikawa} T.,  1995, \mn@doi [\pasp]
  {10.1086/133643}, \href {http://adsabs.harvard.edu/abs/1995PASP..107..945F}
  {107, 945}

\bibitem[\protect\citeauthoryear{{Gregory}, {Scott}, {Douglas}  \&
  {Condon}}{{Gregory} et~al.}{1996}]{Gregory1996ApJS}
{Gregory} P.~C.,  {Scott} W.~K.,  {Douglas} K.,   {Condon} J.~J.,  1996,
  \mn@doi [\apjs] {10.1086/192282}, \href
  {http://cdsads.u-strasbg.fr/cgi-bin/nph-bib_query?bibcode=1996ApJS..103..427G&db_key=AST}
  {103, 427}

\bibitem[\protect\citeauthoryear{{Hovatta} et~al.,}{{Hovatta}
  et~al.}{2014}]{2014MNRAS.439..690H}
{Hovatta} T.,  et~al., 2014, \mn@doi [\mnras] {10.1093/mnras/stt2494}, \href
  {http://adsabs.harvard.edu/abs/2014MNRAS.439..690H} {439, 690}

\bibitem[\protect\citeauthoryear{{Kiehlmann} et~al.,}{{Kiehlmann}
  et~al.}{2016}]{2016arXiv160300249K}
{Kiehlmann} S.,  et~al., 2016, preprint, \href
  {http://adsabs.harvard.edu/abs/2016arXiv160300249K} {} (\mn@eprint {arXiv}
  {1603.00249})

\bibitem[\protect\citeauthoryear{{Kikuchi}, {Mikami}, {Inoue}, {Tabara}  \&
  {Kato}}{{Kikuchi} et~al.}{1988}]{1988A&A...190L...8K}
{Kikuchi} S.,  {Mikami} Y.,  {Inoue} M.,  {Tabara} H.,   {Kato} T.,  1988,
  \aap, \href {http://esoads.eso.org/abs/1988A%26A...190L...8K} {190, L8}

\bibitem[\protect\citeauthoryear{{King} et~al.,}{{King}
  et~al.}{2014}]{2014MNRAS.442.1706K}
{King} O.~G.,  et~al., 2014, \mn@doi [\mnras] {10.1093/mnras/stu176}, \href
  {http://adsabs.harvard.edu/abs/2014MNRAS.442.1706K} {442, 1706}

\bibitem[\protect\citeauthoryear{{Kirhakos}, {Bahcall}, {Schneider}  \&
  {Kristian}}{{Kirhakos} et~al.}{1999}]{Kirhakos99}
{Kirhakos} S.,  {Bahcall} J.~N.,  {Schneider} D.~P.,   {Kristian} J.,  1999,
  \mn@doi [\apj] {10.1086/307430}, \href
  {http://adsabs.harvard.edu/abs/1999ApJ...520...67K} {520, 67}

\bibitem[\protect\citeauthoryear{{Kotilainen}, {Falomo}  \&
  {Scarpa}}{{Kotilainen} et~al.}{1998}]{Kotilainen1998}
{Kotilainen} J.~K.,  {Falomo} R.,   {Scarpa} R.,  1998, \aap, \href
  {http://adsabs.harvard.edu/abs/1998A%26A...332..503K} {332, 503}

\bibitem[\protect\citeauthoryear{{Kotilainen}, {Hyv{\"o}nen}  \&
  {Falomo}}{{Kotilainen} et~al.}{2005}]{Kotilainen2005}
{Kotilainen} J.~K.,  {Hyv{\"o}nen} T.,   {Falomo} R.,  2005, \mn@doi [\aap]
  {10.1051/0004-6361:20042548}, \href
  {http://adsabs.harvard.edu/abs/2005A%26A...440..831K} {440, 831}

\bibitem[\protect\citeauthoryear{{Lavalley}, {Isobe}  \&
  {Feigelson}}{{Lavalley} et~al.}{1992}]{lavalley92}
{Lavalley} M.,  {Isobe} T.,   {Feigelson} E.,  1992, in {Worrall} D.~M.,
  {Biemesderfer} C.,   {Barnes} J.,  eds,  Astronomical Society of the Pacific
  Conference Series Vol. 25, Astronomical Data Analysis Software and Systems I.
  p.~245

\bibitem[\protect\citeauthoryear{{Lister} et~al.,}{{Lister}
  et~al.}{2011}]{2011ApJ...742...27L}
{Lister} M.~L.,  et~al., 2011, \mn@doi [\apj] {10.1088/0004-637X/742/1/27},
  \href {http://adsabs.harvard.edu/abs/2011ApJ...742...27L} {742, 27}

\bibitem[\protect\citeauthoryear{{Lister}, {Aller}, {Aller}, {Hovatta},
  {Max-Moerbeck}, {Readhead}, {Richards}  \& {Ros}}{{Lister}
  et~al.}{2015}]{2015ApJ...810L...9L}
{Lister} M.~L.,  {Aller} M.~F.,  {Aller} H.~D.,  {Hovatta} T.,  {Max-Moerbeck}
  W.,  {Readhead} A.~C.~S.,  {Richards} J.~L.,   {Ros} E.,  2015, \mn@doi
  [\apjl] {10.1088/2041-8205/810/1/L9}, \href
  {http://adsabs.harvard.edu/abs/2015ApJ...810L...9L} {810, L9}

\bibitem[\protect\citeauthoryear{{Marscher} et~al.,}{{Marscher}
  et~al.}{2008}]{2008Natur.452..966M}
{Marscher} A.~P.,  et~al., 2008, \mn@doi [\nat] {10.1038/nature06895}, \href
  {http://adsabs.harvard.edu/abs/2008Natur.452..966M} {452, 966}

\bibitem[\protect\citeauthoryear{{Marscher} et~al.,}{{Marscher}
  et~al.}{2010}]{2010ApJ...710L.126M}
{Marscher} A.~P.,  et~al., 2010, \mn@doi [\apjl]
  {10.1088/2041-8205/710/2/L126}, \href
  {http://adsabs.harvard.edu/abs/2010ApJ...710L.126M} {710, L126}

\bibitem[\protect\citeauthoryear{{Massaro}, {Giommi}, {Leto}, {Marchegiani},
  {Maselli}, {Perri}, {Piranomonte}  \& {Sclavi}}{{Massaro}
  et~al.}{2009}]{2009A&A...495..691M}
{Massaro} E.,  {Giommi} P.,  {Leto} C.,  {Marchegiani} P.,  {Maselli} A.,
  {Perri} M.,  {Piranomonte} S.,   {Sclavi} S.,  2009, \mn@doi [\aap]
  {10.1051/0004-6361:200810161}, \href
  {http://adsabs.harvard.edu/abs/2009A%26A...495..691M} {495, 691}

\bibitem[\protect\citeauthoryear{{Massaro}, {Maselli}, {Leto}, {Marchegiani},
  {Perri}, {Giommi}  \& {Piranomonte}}{{Massaro}
  et~al.}{2015}]{2015ApSS.357...75M}
{Massaro} E.,  {Maselli} A.,  {Leto} C.,  {Marchegiani} P.,  {Perri} M.,
  {Giommi} P.,   {Piranomonte} S.,  2015, \mn@doi [\apss]
  {10.1007/s10509-015-2254-2}, \href
  {http://adsabs.harvard.edu/abs/2015Ap%26SS.357...75M} {357, 75}

\bibitem[\protect\citeauthoryear{{Monet} et~al.,}{{Monet}
  et~al.}{2003}]{2003AJ....125..984M}
{Monet} D.~G.,  et~al., 2003, \mn@doi [\aj] {10.1086/345888}, \href
  {http://adsabs.harvard.edu/abs/2003AJ....125..984M} {125, 984}

\bibitem[\protect\citeauthoryear{{Nilsson}, {Pursimo}, {Takalo},
  {Sillanp{\"a}{\"a}}, {Pietil{\"a}}  \& {Heidt}}{{Nilsson}
  et~al.}{1999}]{nilsson99}
{Nilsson} K.,  {Pursimo} T.,  {Takalo} L.~O.,  {Sillanp{\"a}{\"a}} A.,
  {Pietil{\"a}} H.,   {Heidt} J.,  1999, \mn@doi [\pasp] {10.1086/316446},
  \href {http://adsabs.harvard.edu/abs/1999PASP..111.1223N} {111, 1223}

\bibitem[\protect\citeauthoryear{{Nilsson}, {Pursimo}, {Heidt}, {Takalo},
  {Sillanp{\"a}{\"a}}  \& {Brinkmann}}{{Nilsson}
  et~al.}{2003a}]{2003A&A...400...95N}
{Nilsson} K.,  {Pursimo} T.,  {Heidt} J.,  {Takalo} L.~O.,  {Sillanp{\"a}{\"a}}
  A.,   {Brinkmann} W.,  2003a, \mn@doi [\aap] {10.1051/0004-6361:20021861},
  \href {http://adsabs.harvard.edu/abs/2003A%26A...400...95N} {400, 95}

\bibitem[\protect\citeauthoryear{{Nilsson}, {Pursimo}, {Heidt}, {Takalo},
  {Sillanp{\"a}{\"a}}  \& {Brinkmann}}{{Nilsson} et~al.}{2003b}]{Nilsson03}
{Nilsson} K.,  {Pursimo} T.,  {Heidt} J.,  {Takalo} L.~O.,  {Sillanp{\"a}{\"a}}
  A.,   {Brinkmann} W.,  2003b, \mn@doi [\aap] {10.1051/0004-6361:20021861},
  \href {http://adsabs.harvard.edu/abs/2003A%26A...400...95N} {400, 95}

\bibitem[\protect\citeauthoryear{{Nilsson}, {Pursimo}, {Sillanp{\"a}{\"a}},
  {Takalo}  \& {Lindfors}}{{Nilsson} et~al.}{2008}]{Nilsson08}
{Nilsson} K.,  {Pursimo} T.,  {Sillanp{\"a}{\"a}} A.,  {Takalo} L.~O.,
  {Lindfors} E.,  2008, \mn@doi [\aap] {10.1051/0004-6361:200810310}, \href
  {http://cdsads.u-strasbg.fr/abs/2008A%26A...487L..29N} {487, L29}

\bibitem[\protect\citeauthoryear{{Nilsson}, {Pursimo}, {Villforth}, {Lindfors}
  \& {Takalo}}{{Nilsson} et~al.}{2009}]{nilsson09}
{Nilsson} K.,  {Pursimo} T.,  {Villforth} C.,  {Lindfors} E.,   {Takalo} L.~O.,
   2009, \mn@doi [\aap] {10.1051/0004-6361/200912820}, \href
  {http://adsabs.harvard.edu/abs/2009A%26A...505..601N} {505, 601}

\bibitem[\protect\citeauthoryear{{Nolan} et~al.,}{{Nolan}
  et~al.}{2012}]{2012ApJS..199...31N}
{Nolan} P.~L.,  et~al., 2012, \mn@doi [\apjs] {10.1088/0067-0049/199/2/31},
  \href {http://adsabs.harvard.edu/abs/2012ApJS..199...31N} {199, 31}

\bibitem[\protect\citeauthoryear{{Ofek} et~al.,}{{Ofek}
  et~al.}{2012}]{2012PASP..124..854O}
{Ofek} E.~O.,  et~al., 2012, \mn@doi [\pasp] {10.1086/666978}, \href
  {http://adsabs.harvard.edu/abs/2012PASP..124..854O} {124, 854}

\bibitem[\protect\citeauthoryear{{Pacholczyk}}{{Pacholczyk}}{1970}]{1970ranp.book.....P}
{Pacholczyk} A.~G.,  1970, {Radio astrophysics. Nonthermal processes in
  galactic and extragalactic sources}

\bibitem[\protect\citeauthoryear{{Panopoulou} et~al.,}{{Panopoulou}
  et~al.}{2015}]{2015MNRAS.452..715P}
{Panopoulou} G.,  et~al., 2015, \mn@doi [\mnras] {10.1093/mnras/stv1301}, \href
  {http://adsabs.harvard.edu/abs/2015MNRAS.452..715P} {452, 715}

\bibitem[\protect\citeauthoryear{{Pavlidou} et~al.,}{{Pavlidou}
  et~al.}{2014}]{2014MNRAS.442.1693P}
{Pavlidou} V.,  et~al., 2014, \mn@doi [\mnras] {10.1093/mnras/stu904}, \href
  {http://adsabs.harvard.edu/abs/2014MNRAS.442.1693P} {442, 1693}

\bibitem[\protect\citeauthoryear{{Richards} et~al.,}{{Richards}
  et~al.}{2011}]{2011ApJS..194...29R}
{Richards} J.~L.,  et~al., 2011, \mn@doi [\apjs] {10.1088/0067-0049/194/2/29},
  \href {http://adsabs.harvard.edu/abs/2011ApJS..194...29R} {194, 29}

\bibitem[\protect\citeauthoryear{{Richards}, {Hovatta}, {Max-Moerbeck},
  {Pavlidou}, {Pearson}  \& {Readhead}}{{Richards}
  et~al.}{2014}]{2014MNRAS.438.3058R}
{Richards} J.~L.,  {Hovatta} T.,  {Max-Moerbeck} W.,  {Pavlidou} V.,  {Pearson}
  T.~J.,   {Readhead} A.~C.~S.,  2014, \mn@doi [\mnras]
  {10.1093/mnras/stt2412}, \href
  {http://adsabs.harvard.edu/abs/2014MNRAS.438.3058R} {438, 3058}

\bibitem[\protect\citeauthoryear{{Roy}}{{Roy}}{1995}]{1995PASA...12..273R}
{Roy} A.~L.,  1995, \pasa, \href
  {http://adsabs.harvard.edu/abs/1995PASA...12..273R} {12, 273}

\bibitem[\protect\citeauthoryear{{Saikia} \& {Salter}}{{Saikia} \&
  {Salter}}{1988}]{1988ARA&A..26...93S}
{Saikia} D.~J.,  {Salter} C.~J.,  1988, \mn@doi [\araa]
  {10.1146/annurev.aa.26.090188.000521}, \href
  {http://adsabs.harvard.edu/abs/1988ARA%26A..26...93S} {26, 93}

\bibitem[\protect\citeauthoryear{{Savolainen}, {Homan}, {Hovatta}, {Kadler},
  {Kovalev}, {Lister}, {Ros}  \& {Zensus}}{{Savolainen}
  et~al.}{2010}]{2010A&A...512A..24S}
{Savolainen} T.,  {Homan} D.~C.,  {Hovatta} T.,  {Kadler} M.,  {Kovalev} Y.~Y.,
   {Lister} M.~L.,  {Ros} E.,   {Zensus} J.~A.,  2010, \mn@doi [\aap]
  {10.1051/0004-6361/200913740}, \href
  {http://adsabs.harvard.edu/abs/2010A%26A...512A..24S} {512, A24}

\bibitem[\protect\citeauthoryear{{Sazonov}}{{Sazonov}}{1972}]{1972Ap&SS..19...25S}
{Sazonov} V.~N.,  1972, \mn@doi [\apss] {10.1007/BF00643165}, \href
  {http://adsabs.harvard.edu/abs/1972Ap%26SS..19...25S} {19, 25}

\bibitem[\protect\citeauthoryear{{Scarpa}, {Urry}, {Padovani}, {Calzetti}  \&
  {O'Dowd}}{{Scarpa} et~al.}{2000}]{Scarpa00b}
{Scarpa} R.,  {Urry} C.~M.,  {Padovani} P.,  {Calzetti} D.,   {O'Dowd} M.,
  2000, \mn@doi [\apj] {10.1086/317199}, \href
  {http://adsabs.harvard.edu/abs/2000ApJ...544..258S} {544, 258}

\bibitem[\protect\citeauthoryear{{Shaw} et~al.,}{{Shaw} et~al.}{2013}]{shaw13}
{Shaw} M.~S.,  et~al., 2013, \mn@doi [\apj] {10.1088/0004-637X/764/2/135},
  \href {http://adsabs.harvard.edu/abs/2013ApJ...764..135S} {764, 135}

\bibitem[\protect\citeauthoryear{{Strittmatter}, {Serkowski}, {Carswell},
  {Stein}, {Merrill}  \& {Burbidge}}{{Strittmatter}
  et~al.}{1972}]{1972ApJ...175L...7S}
{Strittmatter} P.~A.,  {Serkowski} K.,  {Carswell} R.,  {Stein} W.~A.,
  {Merrill} K.~M.,   {Burbidge} E.~M.,  1972, \mn@doi [\apjl] {10.1086/180974},
  \href {http://adsabs.harvard.edu/abs/1972ApJ...175L...7S} {175, L7}

\bibitem[\protect\citeauthoryear{{Uemura} et~al.,}{{Uemura}
  et~al.}{2010}]{2010PASJ...62...69U}
{Uemura} M.,  et~al., 2010, \mn@doi [\pasj] {10.1093/pasj/62.1.69}, \href
  {http://adsabs.harvard.edu/abs/2010PASJ...62...69U} {62, 69}

\bibitem[\protect\citeauthoryear{{Urry}, {Scarpa}, {O'Dowd}, {Falomo}, {Pesce}
  \& {Treves}}{{Urry} et~al.}{2000}]{Urry2000}
{Urry} C.~M.,  {Scarpa} R.,  {O'Dowd} M.,  {Falomo} R.,  {Pesce} J.~E.,
  {Treves} A.,  2000, \mn@doi [\apj] {10.1086/308616}, \href
  {http://adsabs.harvard.edu/abs/2000ApJ...532..816U} {532, 816}

\bibitem[\protect\citeauthoryear{{White}, {Becker}, {Helfand}  \&
  {Gregg}}{{White} et~al.}{1997}]{White1997ApJ}
{White} R.~L.,  {Becker} R.~H.,  {Helfand} D.~J.,   {Gregg} M.~D.,  1997,
  \mn@doi [\apj] {10.1086/303564}, \href
  {http://adsabs.harvard.edu/cgi-bin/nph-bib_query?bibcode=1997ApJ...475..479W&db_key=AST}
  {475, 479}

\bibitem[\protect\citeauthoryear{{Wright}, {Griffith}, {Burke}  \&
  {Ekers}}{{Wright} et~al.}{1994}]{Wright1994ApJS}
{Wright} A.~E.,  {Griffith} M.~R.,  {Burke} B.~F.,   {Ekers} R.~D.,  1994,
  \mn@doi [\apjs] {10.1086/191939}, \href
  {http://adsabs.harvard.edu/cgi-bin/nph-bib_query?bibcode=1994ApJS...91..111W&db_key=AST}
  {91, 111}

\bibitem[\protect\citeauthoryear{{Xiong}, {Zhang}, {Bai}  \& {Zhang}}{{Xiong}
  et~al.}{2015}]{2015MNRAS.450.3568X}
{Xiong} D.,  {Zhang} X.,  {Bai} J.,   {Zhang} H.,  2015, \mn@doi [\mnras]
  {10.1093/mnras/stv812}, \href
  {http://adsabs.harvard.edu/abs/2015MNRAS.450.3568X} {450, 3568}

\bibitem[\protect\citeauthoryear{{Yuan}, {Tran}, {Wills}  \& {Wills}}{{Yuan}
  et~al.}{1998}]{1998AAS...19310714Y}
{Yuan} J.,  {Tran} H.~D.,  {Wills} B.~J.,   {Wills} D.,  1998, in American
  Astronomical Society Meeting Abstracts. p. 107.14

\bibitem[\protect\citeauthoryear{{Zhang}, {Chen}  \& {B{\"o}ttcher}}{{Zhang}
  et~al.}{2014}]{2014ApJ...789...66Z}
{Zhang} H.,  {Chen} X.,   {B{\"o}ttcher} M.,  2014, \mn@doi [\apj]
  {10.1088/0004-637X/789/1/66}, \href
  {http://adsabs.harvard.edu/abs/2014ApJ...789...66Z} {789, 66}

\bibitem[\protect\citeauthoryear{{Zhang}, {Chen}, {B{\"o}ttcher}, {Guo}  \&
  {Li}}{{Zhang} et~al.}{2015}]{2015ApJ...804...58Z}
{Zhang} H.,  {Chen} X.,  {B{\"o}ttcher} M.,  {Guo} F.,   {Li} H.,  2015,
  \mn@doi [\apj] {10.1088/0004-637X/804/1/58}, \href
  {http://adsabs.harvard.edu/abs/2015ApJ...804...58Z} {804, 58}

\bibitem[\protect\citeauthoryear{{van Dokkum} \& {Franx}}{{van Dokkum} \&
  {Franx}}{1995}]{1995AJ....110.2027V}
{van Dokkum} P.~G.,  {Franx} M.,  1995, \mn@doi [\aj] {10.1086/117667}, \href
  {http://adsabs.harvard.edu/abs/1995AJ....110.2027V} {110, 2027}

\makeatother
\end{thebibliography}




\appendix

\section{Correcting for host galaxy contribution}                                                                                                
\label{sec:host_contr}

Host galaxy contribution magnitudes as well as limits have been sought in the literature for our all the
sources in our sample. Most of the host galaxy magnitude estimates are obtained by modeling the core and
galaxy emission using a de Vaucouleurs intensity profile integrated to infinity
\citep[e.g.,][]{nilsson99}. Whenever the effective radius of the galaxy was available, we estimated the
contribution of the host galaxy to our magnitude estimates by integrating up to 2.2'', the median aperture
size in our observations using the equations described in \cite{nilsson09}. Host galaxy magnitudes and limit
measured in filters other than in $R$-band we converted between the magnitude systems by using average color
relations for elliptical galaxies from \cite{Kotilainen1998} and \cite{fukugita95}: $R-H = 2.5$, $H-K=0.2$,
and $R-I=0.7$.

In \cite{shaw13}, the absolute magnitude of the host galaxy is estimated from the spectra instead of fitting
images. We convert their absolute magnitudes to apparent magnitudes using the cosmological parameters listed
in their paper. The apparent $R$-band host galaxy magnitudes we use in our analysis are tabulated in
Table~\ref{tab:hosts}. Altogether these were available for 38 objects in our sample 33 of which are in the GL
sample. The magnitude estimates include the correction for the finite aperture size and are not corrected for
Galactic extinction.

For one of our sources, namely RBPLJ1203$+$6031, the host galaxy appears brighter than our mean observed
magnitude. Its host estimate is taken from \cite{shaw13}. It is then possible that the host galaxy estimate as
computed from the spectra have larger systematic uncertainties than the estimates from direct imaging. We also
do not account for the finite aperture size in our observations so that if the host is very extended, it may
be that only a small portion falls within our aperture. The host estimate for RBPLJ1751$+$0939 is taken from
\cite{Scarpa00b} where the imaging was done with an $H$-band filter. It is then possible that the typical
elliptical galaxy color we use to convert between the $H$ and $R$-band filters is not accurate for this
source. For these sources the host contribution correction is omitted.

The correction for the host galaxy contribution has been applied only to the luminosities. Formally, it should
also be applied to the polarization fraction. However, the difference between the corrected and observed
polarization fraction does not exceed the uncertainties in the polarization fraction. Therefore, we omit this
correction.

%
\begin{table*}
 \centering
 \caption{\label{tab:hosts} The host galaxy magnitudes for 33 GL sources. For one of these sources the host
   magnitude estimate $m_\mathrm{host}$ is smaller that the mean magnitude of the target making the host
   contribution removal insensible.}
  \begin{tabular}{llccllccllccllcc}
  \hline
ID     &$\left<R\right>$  &$R_\mathrm{host}$ &Reference &ID  &$\left<R\right>$  &$R_\mathrm{host}$ &Reference &ID  &$\left<R\right>$ &$R_\mathrm{host}$ &Reference \\
       &     &     &          &    &     &     &          &    &    &     &          \\
(RBPL...) &(mag) &(mag) &  &(RBPL...) &(mag) &(mag) &  &(RBPL...)  &(mag)   &(mag)     &          \\
\hline
J0217$+$0837 &15.75 &16.46            &1  &J1217$+$3007  &14.16 &17.49    &5  &J1751$+$0939  &15.90 &16.32 &9 \\      
J0339$-$0146 &16.67 &19.10            &2  &J1229$+$0203  &12.28 &17.25    &2  &J1800$+$7828  &15.46 &17.12            &8 \\   
J0423$-$0120 &17.77 &20.32            &2  &J1248$+$5820  &15.11 &21.20    &4  &J1806$+$6949  &14.43 &16.05            &5 \\   
J0721$+$7120 &14.31 &18.20            &3  &J1256$-$0547  &14.87 &20.00    &6  &J1813$+$3144  &16.38 &18.85            &5 \\   
J0738$+$1742 &15.15 &20.44            &4  &J1427$+$2348  &13.87 &21.00    &4  &J1824$+$5651  &16.19 &20.85            &5 \\   
J0818$+$4222 &17.85 &21.47            &4  &J1512$-$0905  &15.50 &18.70    &2  &J1838$+$4802  &15.56 &19.48            &5 \\  
J0854$+$2006 &14.88 &18.10            &5  &J1555$+$1111  &13.97 &21.60    &4  &J1959$+$6508  &14.03 &16.28            &4 \\   
J0958$+$6533 &16.30 &19.60            &5  &J1642$+$3948  &17.30 &20.22    &7  &J2005$+$7752  &15.95 &20.47            &4 \\   
J1058$+$5628 &15.51 &16.86            &1  &J1653$+$3945  &13.73 &14.85    &5  &J2143$+$1743  &15.65 &18.53            &10\\   
J1132$+$0034 &17.02 &20.45            &1  &J1725$+$1152  &14.45 &21.40    &4  &J2202$+$4216  &13.05 &16.82            &5 \\   
J1203$+$6031 &15.73 &15.34$^\mathrm{a}$ &1  &J1748$+$70 05 &14.90 &18.24    &8  &J2251$+$4030  &16.50 &17.99            &1 \\     
\hline                                                                                           
\end{tabular}                                             
\scriptsize                                               
\begin{flushleft}
$^\mathrm{a}$: $R_\mathrm{host}<\left<R\right>$ \\
References:\\1: \cite{shaw13}, 2: \cite{Kotilainen1998}, 3: \cite{Nilsson08}, 4: \cite{Urry2000}, 5:
\cite{Nilsson03}, 6: \cite{nilsson09}, 7: \cite{Kirhakos99}, 8: \cite{Kotilainen2005}, 9: \cite{Scarpa00b},
10: \cite{Dunlop03}\\
\end{flushleft}
\end{table*}


\bsp	
\label{lastpage}
\end{document}